\documentclass[12pt]{article}

\usepackage{a4wide}
\usepackage[pdftex,usenames,dvipsnames]{color}
\usepackage{graphics}
\usepackage{amsfonts}
\usepackage{cite}
\usepackage[utf8]{inputenc}
\pdfoutput=1

\newcommand{\nit}{\noindent}

\newcommand{\np}{\newpage}
\newcommand{\dsp}{\displaystyle}
\newcommand{\vs}[1]{\vspace{#1 ex}}
\newcommand{\hs}[1]{\hspace{#1 em}}
\newcommand{\bfr}{\begin{flushright}}
\newcommand{\efr}{\end{flushright}}
\newcommand{\bc}{\begin{center}}
\newcommand{\ec}{\end{center}}
\newcommand{\ben}{\begin{enumerate}}
\newcommand{\een}{\end{enumerate}}

\newcommand{\be}{\begin{equation}}
\newcommand{\ee}{\end{equation}}
\newcommand{\ba}{\begin{array}}
\newcommand{\ea}{\end{array}}
\newcommand{\lhs}{\left[}
\newcommand{\rhs}{\right]}
\newcommand{\ct}{\cite}

\newcommand{\ag}{\alpha}
\newcommand{\bg}{\beta}
\newcommand{\gam}{\gamma}
\newcommand{\del}{\delta}

\newcommand{\ve}{\varepsilon}
\newcommand{\zg}{\zeta}
\newcommand{\thg}{\theta}
\newcommand{\kg}{\kappa}
\newcommand{\lb}{\lambda}
\newcommand{\sg}{\sigma}
\newcommand{\rg}{\rho}

\newcommand{\vf}{\varphi}
\newcommand{\og}{\omega}
\newcommand{\Gam}{\Gamma}
\newcommand{\Del}{\Delta}

\newcommand{\Sg}{\Sigma}
\newcommand{\Og}{\Omega}

\newcommand{\lh}{\left(}
\newcommand{\rh}{\right)}

\newcommand{\nb}{\nabla}

\newcommand{\sn}{\mbox{\,sin\,}}
\newcommand{\cs}{\mbox{\,cos\,}}

\newcommand{\ctg}{\mbox{\,cotan\,}}

\newcommand{\der}{\partial}

\begin{document}

\pagestyle{empty}

\bc
{\Large {\bf Spin-induced motion in black hole spacetime}}   \\

\vs{7}

{\large Satish Kumar Saravanan$^{\ast}$}\\
\vs{2}
{International Institute of Physics, Federal University of Rio Grande do Norte, \\
Campus Universitario, Lagoa Nova, Natal-RN 59078-970, Brazil \\
}

\vs{2}
\today
\ec
\vs{5}

\nit
{\footnotesize {\bf Abstract}} \\
Based on the covariant hamiltonian formalism, we study the dynamics of spinning test bodies in the Kerr and Schwarzschild spacetimes. For the first time, we derive the exact solution of circular orbits in the Kerr plane without truncating the spin of the particle or black hole. A large class of noncircular bound orbits has been developed by using the world line perturbation theory. It is found that the spinning body possesses a double frequency, and thus, in addition to the angular shift, the periastron varies radially proportional to the alignment and magnitude of spins. By using the method of stability analysis, we predict the radius of the innermost stable circular orbit (ISCO) as a function of particle and black hole spin. Furthermore, extending the perturbative technique for generic orientation of spin in Schwarzschild spacetime leads to the development of nonplanar bound orbits and the prediction of fully relativistic precessional frequency of the orbital plane.

\vfill

{\footnoterule
\nit
{\footnotesize$^\ast${Email: blackboard.chalkpiece@gmail.com}} \\

\np
\pagestyle{plain}
\pagenumbering{arabic}

\section{Introduction \label{ps1}}
Spin is an indispensable property of astrophysical objects. The imprint of it can be seen in the orbits or gravitational radiation emitted during the coalescence of compact objects. The description of the gravitational two-body problem incorporating the effects of spin has been a subject of theoretical interest \ct{deSitter:1916zza,Thomas:1926dy,Frenkel:1926zz,Mathisson:1937zz,Papapetrou:1951pa,Fock:1954,Dixon:1970zza,Wald:1972sz,Hanson:1974qy} since general relativity's inception. Investigations got intense with the prediction of the first binary pulsar \ct{osti_4215694}, which contains rapidly rotating neutron stars acting as a laboratory for relativistic gravity. Soon after the discovery, a non-geometrical semi-relativistic analysis based on Schwinger's source theory was given \ct{Cho:1975yr}. 

However, the most commonly available explanations are based on the Mathisson - Papapetrou equations \ct{Semerak:1999qc,Schaefer:2004qh,Semerak:2007,Plyatsko:2011gf,Steinhoff:2010zz,Costa:2012cy,Porto:2016pyg}, which account for the internal structure of the extended spinning body. Here, the evolution equations are derived by demanding the covariant conservation of the energy-momentum tensor of matter together with the Einstein field equations. The non-covariant momentum used in this construction is not proportional to the four-velocity. Hence, constraints like Pirani or Tulczyjew specifying some center of mass are additionally needed to define the world line. For a long time, this method was considered as the only general relativistic covariant approach \ct{Karpov:2004su}.

In contrast, a constraint-free covariant formalism describing the spinning particle dynamics has been recently developed \ct{d'Ambrosi:2015gsa}. In this approach, the internal structure of the test body containing spin degrees of freedom is neglected, upon emphasizing the mass dipole moment or Pirani vector cannot be taken to vanish during the evolution. For a given kinetic hamiltonian proportional to the proper four-velocity, the particle obeys the first-order differential equations associated with the world line in which the spin tensor is covariantly constant. Thus, by fixing the initial conditions, the evolution of the system is determined -- completely and uniquely.

Following the latter approach, an analysis of the dynamics of spinning test bodies in the plane of Schwarzschild and Reissner-Nordstr{\o}m backgrounds was established recently \ct{dAmbrosi:2015wqz}. In this paper, we continue to extend this framework for spinning bodies in the test-particle limit ($m<<M$) as follows: In Sec.\ 2, we start describing the equations of motion developed from the particle phase-space for a minimal choice of the kinetic hamiltonian. Then, we collect the conserved quantities resulting from the specific spacetime manifolds and universal ones. The formalism has been applied to the case of the Kerr plane in Sec.\ 3, and the \emph{exact} solution for the circular orbits is found. In Sec.\ 4, a large class of noncircular planar orbits has been developed by using the world line perturbative technique \ct{vanHolten:2016mxa}. It is shown that, in addition to the precession, the radius varies proportionally to the alignment and magnitude of spins. By analyzing the stability of circular orbits, the radius of the innermost stable circular orbit (ISCO) as a function of the spin of the particle and the black hole is given in Sec.\ 5. For generic orientation of spin, the perturbative analysis is applied to the case of nonplanar orbits in Schwarzschild geometry and the precessional frequency of the orbital plane is predicted with an analytical formula in Sec.\ 6. Finally, in Sec.\ 7, we conclude with a summary and future directions. Dimensions of physical quantities and mathematical details are collected in appendixes.

\section{Spinning particles: hamiltonian formulation and conserved quantities \label{ps2}}
The equations of motion for a structure-less point particle with internal angular momentum (spin) tracing the world line $x^{\mu}(\tau)$ in curved space-time in an effective world-line formalism read:
\be
\ba{l}
\dsp{ D_{\tau} u^{\mu}} = \dsp{\dot{u}^{\mu} + \Gam^{\mu}_{\lb\nu} u^{\lb} u^{\nu}} =  \dsp{ \frac{1}{2 m}\, \Sg^{\kg\lb} R_{\kg\lb\;\,\nu}^{\;\;\;\,\mu} u^{\nu}}, \\
\\
\dsp{D_{\tau} \Sg^{\mu\nu}}  = \dsp{\dot{\Sg}^{\mu\nu} + u^{\lb} \lh \Gam^{\mu}_{\lb\kg} \Sg^{\kg\nu} +  \Gam^{\nu}_{\lb\kg} \Sg^{\mu\kg} \rh = 0}.
\ea
\label{2.1}
\ee
$D_\tau$ denotes a covariant derivative, and the overdot denotes an ordinary derivative w.r.t proper time, $\tau$, and $m$ is the spinning particle's (constant) mass. Then $u^\mu = \dot{x}^{\mu}$ is the proper four-velocity obeying the condition: $u_\mu u^{\mu} = -1$ for time-like unit vector. The anti-symmetric spin tensor $\Sg^{\mu\nu}$ unifies the internal angular momentum pseudo-vector $S^{\mu}$  and the mass dipole moment $Z^{\mu}$:
\be
S_{\mu} = \frac{1}{2}\, \sqrt{-g}\, \ve_{\mu\nu\kg\lb} u^{\nu} \Sg^{\kg\lb}, \hs{1}
Z_{\mu} = \Sg_{\mu\nu} u^{\nu}.
\label{2.2}
\ee
Both vectors are space-like:
\be
S_{\mu}u^{\mu} = Z_{\mu}u^{\mu} = 0,
\ee
as required to match the number of independent components of  $\Sg^{\mu\nu}$. Equivalently, the relations (\ref{2.2}) can also be inverted as 
\be
\Sg^{\mu\nu} = -\frac{1}{\sqrt{-g}}\ve^{\mu\nu\kg\lb}u_{\kg}S_{\lb} + u^{\mu} Z^{\nu} - u^{\nu} Z^{\mu}. 
\ee
When the particle has vanishing spin ($\Sg^{\mu\nu} =0$), it moves along the geodesics as equations (\ref{2.1}) reduce to geodesic equations of motion. Whereas in the case of spinning particles, for the given initial conditions the combined solution of equations (\ref{2.1}) traces the curve in spacetime i.e., world line. The spin tensor coupling to the spacetime curvature of the external geometry  implies the gravitational Lorentz force in the first equation. The second equation denotes the spin tensor is covariantly constant along the world line. These equations can be obtained either in a hamiltonian formulation or from local energy-momentum conservation \ct{vanHolten:2015vfa}. 

In the hamiltonian approach, the equations of motion can be derived by choosing a specific hamiltonian and a set of classical Dirac-Poisson brackets. For a massive free-spinning particle in the absence of Stern-Gerlach forces and external fields, the hamiltonian is 
\be
H_0 = \frac{1}{2m}\, g^{\mu\nu}(x) \pi_{\mu} \pi_{\nu},   \hs{2}  \pi_{\mu} = m g_{\mu\nu} u^{\nu},
\label{2.3}
\ee
where $\pi_{\mu}$ is identified as the covariant or kinetic momentum proportional to the proper-four velocity. The equations of motion for	 generalised hamiltonians incorporating complicated interactions have been derived  in our earlier works \ct{dAmbrosi:2015wqz}. However, in this paper we study the dynamics generated by the minimal hamiltonian only. 

A spinning particle's phase-space is finite-dimensional and consists of the position $x^{\mu}$, covariant momentum $\pi_{\mu}$ and spin tensor $\Sg^{\mu\nu}$. These dynamical degrees satisfy the covariant Dirac-Poisson brackets: 
\be
\ba{l}
\left\{ x^{\mu}, \pi_{\nu} \right\} = \del^{\mu}_{\nu}, \hs{2}
\left\{ \pi_{\mu}, \pi_{\nu} \right\} = \frac{1}{2}\, \Sg^{\kg\lb} R_{\kg\lb\mu\nu}, \\
 \\
\left\{ \Sg^{\mu\nu}, \pi_{\lb} \right\} = \Gam_{\lb\kg}^{\;\;\;\mu}\, \Sg^{\nu\kg} - \Gam_{\lb\kg}^{\;\;\;\nu}\, \Sg^{\mu\kg}, \\
 \\
\left\{ \Sg^{\mu\nu}, \Sg^{\kg\lb} \right\} = g^{\mu\kg} \Sg^{\nu\lb} - g^{\mu\lb} \Sg^{\nu\kg} - g^{\nu\kg} \Sg^{\mu\lb}
 + g^{\nu\lb} \Sg^{\mu\kg}.
\ea
\label{2.4}
\ee
It is formulated by demanding the phase-space variables obey the Jacobi identities for cyclic triple brackets
\be
 \left\{x^{\mu}, \left\{\pi_{\nu},\Sg^{\kg\lb}\right\}\right\} + \left\{ \pi_{\nu}, \left\{\Sg^{\kg\lb},x^{\mu}\right\}\right\} + \left\{\Sg^{\kg\lb} \left\{x^{\mu}, \pi_{\nu}\right\}\right\} = 0,
\ee
and the Casimir invariant for the total spin $I$,
\be
\left\{I, x^{\mu}  \right\}  =  \left\{I, \pi_{\mu}  \right\} =  \left\{I, \Sg^{\mu\nu}  \right\} =0,  \hs{2}  I = \frac{1}{2}\Sg_{\mu\nu}\Sg^{\mu\nu}. 
\ee
Since the above conditions are satisfied without specifying the choice of the hamiltonian, the brackets are closed and model-independent. 

Unlike the traditional approaches \ct{Mathisson:1937zz,Papapetrou:1951pa,Dixon:1970zza,Pirani:1956tn}, here the system consists of first-order differential equations. Therefore, for a given hamiltonian  the evolution of the system is uniquely determined by the initial conditions.  Thus, we do not need any additional constraints. The equations  (\ref{2.1}) that define the world-line by parallel transport of the spin-tensor while allowing for a non-vanishing mass dipole $Z$ provide a simpler representation of the orbit than approaches focussing on the motion of a (non-unique) center of mass.  Individually, the mass dipole $Z$ and spin proper $S$ manifest as proper time evolution equations in our formalism:
\be
\ba{l}
\dsp{D_{\tau} S^{\mu} = \frac{1}{4m \sqrt{-g}}\, \ve^{\mu\nu\kg\lb} \Sg_{\kg\lb} \Sg^{\ag\bg} R_{\ag\bg\nu\rho} u^{\rho}, }\\
 \\
\dsp{D_{\tau} Z^{\mu} =  \frac{1}{2 m}\, \Sg^{\mu\nu} \Sg^{\ag\bg} R_{\ag\bg\nu\rho} u^{\rho}. }
\ea
\label{2.5}
\ee
It doesn't mean that the two formulations are contradictory. Both are in agreement with the limit of linearized spin.

This formulation is also convenient for deriving constants of motion, which commute with the hamiltonian in the sense of the brackets. In particular, the usual relations between constants of motion and Killing vectors of the space-time manifold are generalized for spinning particles as
\be
\left\{J, H_{0}\right\} = 0 \Rightarrow  \hs{2} J = \ag^{\mu} \pi_{\mu} + \frac{1}{2}\, \bg_{\mu\nu} \Sg^{\mu\nu}.
\label{2.6}
\ee
$J$ is a constant of motion if $\ag^{\mu}(x)$ is a Killing vector field and $\bg_{\mu\nu}$ its normalized curl:
\be
\nb_{\mu} \ag_{\nu} + \nb_{\nu} \ag_{\mu} = 0, \hs{2}   \nb_{\mu} \ag_{\nu} - \nb_{\nu} \ag_{\mu}  = 2 \bg_{\mu\nu},
\ee
with an identity,
\be
\nb_{\lb} \bg_{\mu\nu} = R_{\mu\nu\lb\kg}\ag^{\kg}.
\ee
Additionally, universal constants of motion are provided by the hamiltonian, the total spin and the pseudo-scalar spin-dipole product:
\be
\ba{l}
\dsp{H_0 = - \frac{m}{2} \Rightarrow g_{\mu\nu}u^{\mu}u^{\nu} = -1}, \\
\\
\dsp{I = \frac{1}{2}\, g_{\mu\kg} g_{\nu\lb} \Sg^{\mu\nu} \Sg^{\kg\lb} = S \cdot S + Z \cdot Z,} \\
 \\
 \dsp{D = \frac{1}{8}\, \sqrt{-g}\, \ve_{\mu\nu\kg\lb} \Sg^{\mu\nu} \Sg^{\kg\lb} = S \cdot Z}.
\ea
\label{2.7}
\ee
These conserved quantities will be essential in solving the system in specific spacetime geometries.

\section{The exact solution of circular orbits in Kerr spacetime}
The Kerr metric \ct{Kerr:1963ud} gives the rotating vacuum solution of Einstein's field equation $R_{\mu\nu} = 0$, which is written in \emph{Boyer-Lindquist coordinates} as
\be 
\ba{lll}
\dsp{ ds^2 } & = & - \dsp{\lh 1 - \frac{2Mr}{\rho^2}\rh dt^2 -\frac{4a M r \, \sn^{2}\thg}{\rho^2} d\vf\,dt + \frac{\rho^2}{\Del^2} dr^2 + \rho^2 d\thg^2 }\\
 & & \\
 & & + \dsp{\lh r^2 + a^2 + \frac{2a^2 M r  \, \sn^{2}\thg}{\rho^2} \rh \sn^{2}\thg \, d\vf^2},
\ea
\label{3.1}
\ee
where $r$ is the radial coordinate,  $M$ and $a$ are mass and Kerr parameters, and we label
\be
 \rho^2 \equiv r^2 + a^2 \cs^2\thg, \qquad \Del^2 \equiv r^2 - 2Mr +a^2.
\ee 
The stationary and axisymmetric properties of the above line element admit two killing vector fields, $\ag^{\mu}_t = (1, 0, 0, 0)$ and $\ag^{\mu}_\vf = (0, 0, 0, 1)$ in the geometry. As a result, the generalized conservation laws for spinning particles (\ref{2.6}) manifest as kinetic energy $E$ conservation:
\be 
\ba{lll}
\dsp{\ve} = \dsp{\frac{E}{m}} & = & \dsp { \lh 1 - \frac{2M}{r} \rh u^{t} + \frac{2a  M}{r} u^{\vf} - \frac{M}{m r^2} \Sg^{tr} - \frac{aM}{mr^2} \Sg^{r\vf},} 
 \label{3.2}
\ea
\ee
and  total angular momentum  $J$ conservation:
\be 
\ba{lll}
\dsp{\eta} = \dsp{\frac{J}{m}} & = & \dsp{ - \frac{2a M}{r} u^{t} +   \lh r^2 + a^2 \lh 1+ \frac{2M}{r} \rh \rh u^{\vf} }\\
 & & \\
 & & \dsp{- \frac{a M}{m r^2} \Sg^{tr} + \frac{1}{m r^2}  \lh r^3 - a^2 M \rh \Sg^{r \vf},}
 \label{3.3}
\ea
\ee
on the plane. The direction of the total angular momentum has been chosen as the $z$ axis, i.e., ${\bf{J}} = (0, 0, J)$.  Where $\ve$ and $\eta$ are the total energy and total angular momentum per unit mass of the particle, respectively. Analogous to the case of test mass, we identify (\ref{3.3}) $\eta$ as sum of the orbital $\ell$ and spin angular momentum $\sg$ per unit mass of the particle: $\eta = \ell+ \sg$. This implies, when the spin of the particle vanishes, these conserved quantities reduce to the case of test mass in Kerr. Also it reduce to the spinning particles in Schwarzschild for vanishing black hole spin.

Now, we develop the exact circular orbit solutions for spinning particles in the Kerr field. The calculations are performed with a complete spin of the particle and for full metric. 

For the circularly orbiting particle in the equatorial plane, $\thg = \pi/2$, and with radius $r = R$, the velocity components are
\be
u^{\mu} = (u^{t}, 0, 0, u^{\vf}). 
\label{3.01}
\ee
The rotationally symmetric axis ($\vf$) of angular momentum and time ($t$) independence of the metric denotes the angular velocity $u^{\vf}$ and the gravitational time dilation $u^{t}$ are constants w.r.t proper time. 

Under these conditions, it is straightforward to solve the system (\ref{2.1}) of ten coupled first order equations for orbital and spin components. The orbital equations for $u^t$ and $u^\vf$ 
\be
\ba{l}
\dsp{\dot{u}^{t} = \frac{M}{mR^4}\lh R^3 + a^2 (R+2M) \rh\Sg^{t\vf}u^{\vf}, }\\
 \\
\dsp{\dot{u}^{\vf} = \frac{M}{mR^4}\lh R-2M \rh\Sg^{t\vf}u^{t}, }
\ea
\label{3.11}
\ee
implies $\Sg^{t\vf} = 0$. This in turn predicts that the spin components $\Sg^{tr}$ and $\Sg^{r\vf}$ are constants from the following spin equations:
\be
\ba{l}
\dsp{\dot{\Sg}^{tr} + \frac{\Del^2}{R^4} \lhs -aM u^t + \lh a^2 M - R^3 \rh u^{\vf}  \rhs \Sg^{t\vf} = 0,}\\
 \\
\dsp{\dot{\Sg}^{r\vf} + \frac{M \Del^2}{R^4} \lh u^t -a u^{\vf}  \rh \Sg^{t\vf} = 0.}
\ea
\label{3.12}
\ee
And since the rate of change of  $\Sg^{t\vf}$ must also vanish, the spin equation for $\Sg^{t\vf}$ leaves a constraint between $\Sg^{tr}$ and $\Sg^{r\vf}$ as
\be
\lhs a M u^t + \lh R^2 (R-2M) - a^2 M \rh u^{\vf}  \rhs \Sg^{tr} = - M \lhs (R^2 + a^2) u^t  - a (3R^2 + a^2) u^{\vf}\rhs \Sg^{r\vf}.
\label{3.13}
\ee
Then the orbital equation for $u^{\thg}$ 
\be 
\ba{lll}
&  &  \dsp{ \Sg^{t\thg} \lhs \lh R(R-2M) + 3a^2\rh u^t - a \lh R(3R-2M) + 3a^2 \rh u^{\vf}  \rhs }\\
 & & \\
 & & \dsp{ + \Sg^{\thg\vf} \lhs - \lh 2R^4 + a^2 R (5R - 2M) + 3a^4 \rh u^{\vf}  + a \lh R(3R-2M) + 3a^2 \rh u^{t} \rhs= 0,}
\ea
\label{3.14}
\ee
and the remaining three spin equations for $\thg$-components  
\be
\dot{\Sg}^{t\thg} + \frac{M}{\Del^2 R^2} \lhs (R^2 +a^2) u^t - a (3R^2 +a^2) u^{\vf}  \rhs \Sg^{r \thg} = 0,
\label{rthg}
\ee
\be
\dot{\Sg}^{\thg\vf} - \frac{1}{\Del^2 R^2} \lhs a M u^t + \lh R^2 (R-2M) - a^2 M  \rh u^{\vf}  \rhs \Sg^{r \thg} = 0,
\label{thgvf}
\ee
\be
\dot{\Sg}^{r\thg} + \frac{M \Del^2}{R^4} \lh u^t -a u^{\vf}  \rh \Sg^{t\thg} +  \frac{\Del^2}{R^4} \lhs -aM u^t + \lh a^2 M - R^3 \rh u^{\vf}  \rhs \Sg^{\vf\thg}  = 0,
\label{drthg}
\ee
leaves three unknown quantities: $\Sg^{t\thg}$, $\Sg^{\thg\vf}$, and $\Sg^{r\thg}$. By the method of elimination, it is simple to solve these four equations for three unknowns, and we obtain 
\be
\Sg^{r\thg} = \Sg^{t\thg} = \Sg^{\thg\vf} = 0.
\label{3.16}
\ee
This implies that for spinning particles in equatorial circular orbits, there is no motion in the direction perpendicular to the plane. Because the only remaining non-zero spin components, $\Sg^{tr}$ and $\Sg^{r\vf}$, are constants, the direction and magnitude of the orbital and spin angular momentum must be identically conserved. As a result,  the orbital and spin angular momentum must align either parallel or anti-parallel.  

Finally, we are left with the orbital equation for $u^r$
\be 
\ba{lll}
&& \hs{-1.2} \dsp {\Del^2 \bigg[ Mu^{t\,2} - 2 a M u^t u^{\vf} + (a^2 M - R^3) u^{\vf\,2} \bigg]}   \\
&& \\
 && \hs {1} =  \dsp{\frac{M}{mR} \bigg[ \Sg^{tr}  \bigg( - \lh 2R (R-2M) + 3a^2 \rh u^t + a \lh  R (3R - 4M) + 3a^2 \rh u^{\vf}   \bigg) }\\
 & & \\
 & & \hs {3.8} \dsp{+  \Sg^{r\vf}  \bigg(  \lh R^4 + 4 a^2 R (R-M) +  3a^4\rh u^\vf - a \lh  R (3R - 4M) + 3a^2 \rh u^{t}   \bigg) \bigg]},
\ea
\label{3.17}
\ee	 
which is further reduced into a quadratic form in the azimuthal frequency $\Og_{\vf} (=u^{\vf}/u^{t})$ by incorporating the constraint (\ref{3.13}) between the two non-zero spin components, and by using the conserved quantities $(\ve, \eta)$:
\be
X \Og_{\vf}^2 + Y \Og_{\vf} + Z =0,
\label{3.18}
\ee
where the coefficients $(X, Y, Z)$ are expressed in terms of constants $(R, \sg, a)$, and given in Appendix B. Thus, we write the solution to circular orbits as
\be
\Og_{\vf} = \frac{-Y \pm \sqrt{Y^2 - 4 X Z}}{2X}.
\label{3.19}
\ee
The sign $\pm$ stand for the alignment of spin parallel or anti-parallel w.r.t orbital angular momentum. The Kerr parameter $a$ take positive and negative values for co-rotating and counter-rotating orbits, respectively. Then, it is straight forward to establish the gravitational time dilation $u^t$ by using $ \Og_{\vf}$ in the Hamiltonian constants of motion (\ref{2.7})
\be
u^{t} = \sqrt{\frac{R}{ (R-2M) + 4 a M \Og_{\vf} - (R^3+ a^2 (R+2M))\,  \Og_{\vf}^2}},
\label{3.20}
\ee
and in turn we get the angular velocity $u^{\vf}$:
\be
u^{\vf} = \Og_{\vf} \sqrt{\frac{R}{ (R-2M) + 4 a M \Og_{\vf} - (R^3+ a^2 (R+2M))\,  \Og_{\vf}^2}}. 
\ee
And then the total spin $I$ is written in terms of the above determined quantities as
\be 
\ba{lll}
\dsp{I}  & = & \dsp{ \frac{m^2}{R M^2 \Del^2} \Bigg\{-(R-2M) \bigg[ \lhs (R^2(R-2M)-a^2 M) + a M (3R^2 + a^2) \Og_{\vf}  \rhs u^t }\\
 & & \hs{11} - (R^3-a^2 M)\ve - a M \eta \bigg]^2  \\
 & & \dsp{ \hs{4.7} + (R^3+a^2(R+2M))\bigg[ \lhs a- (R^2+a^2) \Og_{\vf}\rhs u^{t} - a \ve + \eta    \bigg]^2  \Bigg\}}.
 \label{}
\ea
\ee
Finally, in circular orbits, the pseudo-scalar $D$ satisfies the condition:
\be
D = S \cdot Z = 0.
\ee
We complete this section by retrieving the special cases: \\

I. When black hole spin ($a=0$) vanishes, we obtain the frequency of spinning particles in Schwarzschild
\be
\Og_{\vf} = \pm \sqrt{{\frac{M}{R^3}} \lh {\frac{ \eta (R-2M) - R \sg }{ \eta (R-2M) - \sg (R-3M) }}\rh }.
\label{}
\ee
And thus the time dilation $u^{t}$ and the angular velocity $u^{\vf}$ are 
\be
u^{t} =  \sqrt{\frac{R \lhs \eta (R-2M) - \sg (R-3M) \rhs}{\lhs  \eta (R-2M)(R-3M) + \sg \lh M R - (R-2M)(R-3M) \rh \rhs}},
\label{}
\ee
\be
u^{\vf} = \pm \sqrt{\frac{M}{R^2} \frac{\lhs \eta (R-2M) - R \sg \rhs}{\lhs  \eta (R-2M)(R-3M) + \sg \lh M R - (R-2M)(R-3M) \rh \rhs}}.
\label{}
\ee
Here $\pm$ denotes the parallel and anti-parallel alignment of particle's spin w.r.t orbital angular momentum.  Whereas, we have used the the conservation laws $(\ref{3.2}, \ref{3.3})$, the spin constraint $(\ref{3.13})$, and the Hamiltonian constraint $(\ref{3.20})$ to eliminate $\ve$. Evidently, $u^{t}$ and $u^{\vf}$ satisfy as solutions of the circular orbits equation presented earlier \ct{dAmbrosi:2015wqz}.\\

II. For the particle's vanishing spin ($\sg=0$), we get the frequency of the test mass in Kerr
\be
\Og_\vf = \frac{1}{a \pm \sqrt{\frac{R^3}{M}} }.
\label{}
\ee
Where $\pm$ indicates the co-rotation and counter-rotation of the particle, and  the time dilation $u^{t}$ and the angular velocity $u^{\vf}$ reduce to
\be
u^{t} =  \frac{\sqrt{M}\lh a\pm \sqrt{\frac{R^3}{M}} \rh}{\sqrt{R \lh R^2 - 3M R \pm 2a \sqrt{M R}\rh}},      \qquad   u^{\vf} =  \frac{\sqrt{M}}{\sqrt{R \lh R^2 - 3M R \pm 2a \sqrt{M R}\rh}}.
\ee

III. Finally, when the spin of both the objects in the binary vanishes: $(\sg =0)$ and $(a=0)$, it reduces to test mass in Schwarzschild 
\be
\Og_{\vf} = \sqrt{ \frac{M}{R^3}},  \qquad     u^{t} =   \sqrt{\frac{R}{R-3M}},      \qquad   u^{\vf} =   \sqrt{\frac{M}{ R^2 (R-3M)}}.
\ee

\section{Plane noncircular orbits and periastron precession\label{npco}}
The exact solution (\ref{3.19}) to the relativistic circular orbit found in the previous section can be used as a reference orbit to generate completely relativistic noncircular bound orbits. Such a perturbative technique has been analytically developed for particles carrying spin and applied to the planar case of spherically symmetric spacetimes \ct{dAmbrosi:2015wqz}. Here we generalize the analysis to the equatorial plane in the Kerr field.

Suppose in a given spacetime $g_{\mu\nu}$: let $(x^{\mu}, u^{\mu}, \Sg^{\mu\nu})$ and $(\bar{x}^{\mu}, \bar{u}^{\mu}, \bar{\Sg}^{\mu\nu})$ are the worldline parameters of the referance orbit and the infinitesimally deviated orbit. Then the proper time evolution equations for the infinitesimal deviations $(\del x^{\mu}, \del u^{\mu}, \del\Sg^{\mu\nu})$ are computed by requiring that the original and perturbed orbit satisfy the equations of motion (\ref{2.1}). The first order worldline deviations are then given by
\be
\ba{l}
\dsp{ \frac{d^2 \del x^{\mu}}{d\tau^2} + 2 u^{\lb} \Gam_{\lb\nu}^{\;\;\;\mu} \frac{d\del x^{\nu}}{d\tau} + 
 u^{\kg} u^{\lb} \der_{\nu} \Gam_{\kg\lb}^{\;\;\;\mu} \del x^{\nu} }\\
 \\
\dsp{ \hs{2} =\, \frac{1}{2m} \left[ \Sg^{\rg\sg} R_{\rg\sg\;\,\nu}^{\;\;\;\,\mu}\, \frac{d\del x^{\nu}}{d\tau} +
  \Sg^{\rg\sg} \der_{\nu} R_{\rg\sg\;\,\kg}^{\;\;\;\,\mu}\, u^{\kg} \del x^{\nu} + 
  \del \Sg^{\rg\sg} R_{\rg\sg\;\,\nu}^{\;\;\;\,\mu} u^{\nu} \right], }
\ea
\label{ncorbeqn}
\ee
\be
\ba{l} 
\dsp{ \frac{d \del \Sg^{\mu\nu}}{d\tau} + u^{\lb} \Gam_{\lb\kg}^{\;\;\;\mu}\, \del \Sg^{\kg\nu} 
 + u^{\lb} \Gam_{\lb\kg}^{\;\;\;\nu}\, \del \Sg^{\mu\kg} }\\
 \\
\dsp{ \hs{3.3} =\, \lh \Gam_{\lb\kg}^{\;\;\;\mu} \Sg^{\nu\kg} - \Gam_{\lb\kg}^{\;\;\;\nu} \Sg^{\mu\kg} \rh 
  \frac{d\del x^{\lb}}{d\tau} + \lh \der_{\lb} \Gam_{\rg\kg}^{\;\;\;\mu} \Sg^{\nu\kg} -  
          \der_{\lb} \Gam_{\rg\kg}^{\;\;\;\nu} \Sg^{\mu\kg} \rh u^{\rg} \del x^{\lb}. }
\ea
\label{ncspineqn}
\ee
The equivalent covariant expressions are given in the above-mentioned reference. Following a similar line of argument, the higher-order deviations can be derived as given for test particles \ct{Koekoek:2010pv,Koekoek:2011mm,Kerner:2001cw,Puetzfeld:2015uxi}. 

On solving this coupled system of differential equations we establish the world-line parameters as linear functions of proper time with constant coefficients. As we confine our analysis to the plane, the variations in the $\theta$ direction $(\del\thg, \del\Sg^{r\thg}, \del\Sg^{t\thg}, \del\Sg^{\thg\vf})$ vanishes. That is, the condition (\ref{3.16}) holds true for planar orbits. Thus the system is left with three equations for orbital parameters $\del x^{\mu} = (\del t, \del r, \del \vf)$, and three equations for spin degrees of freedom $\del \Sg^{\mu\nu} = (\del\Sg^{tr}, \del\Sg^{r\vf}, \del\Sg^{t\vf})$. Further, by using the conservation laws, the deviations for $\del\Sg^{tr}$ and  $\del\Sg^{r\vf}$ are translated to $\del\ve$, the change in energy, and $\del\eta$, the change in total angular momentum, and to the coordinate variations. Therefore the system of 10 equations reduce to 4, and in compact form we write 
\be
\ba{l}
\lh \ba{cccc}
 \frac{d^2}{d\tau^2} & 0 & \ag \frac{d}{d\tau} & \bg \\
  & & & \\
  0 & \frac{d^2}{d\tau^2} & \gam \frac{d}{d\tau} & \zg \\
  & & & \\
  \kg \frac{d}{d\tau} & \lb \frac{d}{d\tau} & \frac{d^2}{d\tau^2} + \mu & 0 \\
  & & & \\
  \nu \frac{d}{d\tau} & \sg \frac{d}{d\tau} & \chi & \frac{d}{d\tau}
 \ea \rh \lh \ba{c} \del t \\ \\ \del \vf \\ \\ \del r \\ \\ \del \Sg^{t\vf} \ea \rh 
  = \lh \ba{c} 0 \\ \\  0 \\ \\ e \del \eta + f \del \ve \\  \\ g \del \eta + h \del \ve \ea \rh  ,
\ea
\label{mat}
\ee
where the coefficients are evaluated at the circular reference orbit, and collected in the Appendix C. 
								
These inhomogeneous eigenfunctions can be simplified for the purpose of constructing noncircular orbits with the same energy $\ve$ and total angular momentum $\eta$ as the reference orbit. In such cases, the change in energy and total angular momentum must vanish: $\del\ve = \del\eta = 0$. As a result, the system will reduce to a set of homogeneous linear differential equations for the deviation vectors as functions of time $\tau$. Then it is trivial to obtain the characteristic equation for the periodic eigen functions,

\be
\og^3 \lh \og^4 - A \og^2 + B \rh = 0, 
\label{nc1}
\ee
where
\be
\ba{l}
A = \mu - \ag \kg - \bg \nu - \gam \lb - \zg \sg, \\
 \\
B = \bg \lh \kg \chi - \mu\nu + \gam (\lb \nu - \kg \sg) \rh + \zg \lh \lb \chi - \mu \sg - \ag (\lb \nu - \kg \sg) \rh.
\ea
\label{nc2}
\ee
In analogy with the motion of test mass, there are three 0-modes for secular solutions considering the orbital degrees of freedom \ct{Koekoek:2010pv,Koekoek:2011mm}. But since the newly predicted eccentric orbit and the reference circular orbit have the same $\ve$ and $\eta$, these secular solutions can be eliminated. Hence, we are left with two pairs of non-trivial periodic solutions with angular frequencies
\be
\og^2_{\pm} = \frac{1}{2} \lh A \pm \sqrt{A^2 - 4B} \rh.
\label{nc3}
\ee
Thus, we establish the complete first order world line solutions for the orbital and spin degrees of freedom in the equatorial plane as
\be
\ba{l}
t(\tau) = u^t \tau + n^t_+ \sin \og_+ (\tau - \tau_+) + n^t_- \sin \og_- (\tau - \tau_-), \\
 \\
\vf(\tau) = u^{\vf} \tau + n^{\vf}_+  \sin \og_+ (\tau - \tau_+) + n^{\vf}_- \sin \og_- (\tau - \tau_-), \\ 
 \\
r(\tau) = R + n^r_+ \cos \og_+ (\tau - \tau_+) + n^r_- \cos \og_- (\tau - \tau_-),\\
\\
\Sg^{t\vf}(\tau) =  N^{t\vf}_{+} \sin \og_+ (\tau - \tau_+) +  N^{t\vf}_{-} \sin \og_- (\tau - \tau_-), \\
 \\ 
 \Sg^{tr}(\tau) = \Sg^{tr}_{0} + N^{tr}_{+} \cos \og_+ (\tau - \tau_+)  + N^{tr}_{-} \cos \og_- (\tau - \tau_-),  \\
 \\
 \Sg^{r\vf}(\tau) = \Sg^{r\vf}_{0} + N^{r\vf}_{+} \cos \og_+ (\tau - \tau_+)  + N^{r\vf}_{-} \cos \og_- (\tau - \tau_-).  \\
\ea
\label{nc4}    
\ee
Where, $\Sg^{tr}_{0}$ and $\Sg^{r\vf}_{0}$ are the spin tensor components obeying the circular orbits (\ref{3.19}). The amplitudes $n^{\alpha}_{\pm}$ and $N^{\alpha\beta}_{\pm}$ corresponds to orbital and spin deviations respectively, and are expressed in terms of circular orbit constants with $C_{\pm}$ being the common normalization factor:
\be
\ba{lll}
\dsp{n_{\pm}^t} &=& \dsp{C_{\pm} \lhs \lb (\beta \gamma - \ag \zeta) + \beta (\omega_{\pm}^2 - \mu) \rhs}, \\
&& \\
\dsp{n_{\pm}^\vf} &=& \dsp{C_{\pm}  \lhs - \kg (\beta \gamma - \ag \zeta) + \zeta (\omega_{\pm}^2 - \mu) \rhs}, \\
&& \\
\dsp{n_{\pm}^r} &=& \dsp{C_{\pm} \omega_{\pm} (\beta \kg + \zeta \lb)}, \\
&& \\
\dsp{N^{t \vf}_{\pm}} & = & \dsp{C_{\pm} \omega_{\pm}^2 (\omega_{\pm}^2 - \mu + \ag \kg + \gamma \lb)}, \\
&& \\
\dsp{N^{tr}_{\pm}}&=&\dsp{\frac{m}{M R^2} \big[ R(R^2(R-2M) - a^2M) \,\og_{\pm} n^{t}_{\pm}  +a M R (3R^2+a^2) \,\og_{\pm} n^{\vf}_{\pm} } \\
&&\\
&&\dsp{\hs{0.5}  +\lh (2R^2 (R-M)+a^2M) u^t + aM (3R^2-a^2) u^{\vf} - (2R^3+a^2M) \ve + aM \eta \rh n^{r}_{\pm} \big]}, \\
&& \\
\dsp{N^{r\vf}_{\pm}} &=& \dsp{\frac{m}{R^2}\lhs a R\, \og_{\pm} n^{t}_{\pm}  - R (R^2+a^2)\,\og_{\pm} n^{\vf}_{\pm} - \lh a u^t + (R^2-a^2) u^{\vf} - a \ve +\eta \rh  n^{r}_{\pm} \rhs}. 
\label{nc5}    
\ea
\ee 
Because the system has two fundamental frequencies, $\og_\pm$, the periastron and apastron will have different minima and maxima, which are determined by the integration constants, $\tau_\pm$. Such an effect due to spin-curvature coupling has been established for spinning bodies in the Reissner-Nordstr{\o}m geometry \ct{dAmbrosi:2015wqz}. Whereas, in Kerr spacetime, because of additional spin-spin interactions, it also depends on the rotational parameter $a$. Hence, we demonstrate this effect with two examples. In Fig.\ 1. we show, a particle in the counter-rotating orbit with Kerr parameter $a=-M$ and prograde spin $\sg\simeq0.073M$ will deviate radially from the reference circular orbit $R=10M$ as a function of proper time. 
\bc
\scalebox{0.58}{\includegraphics{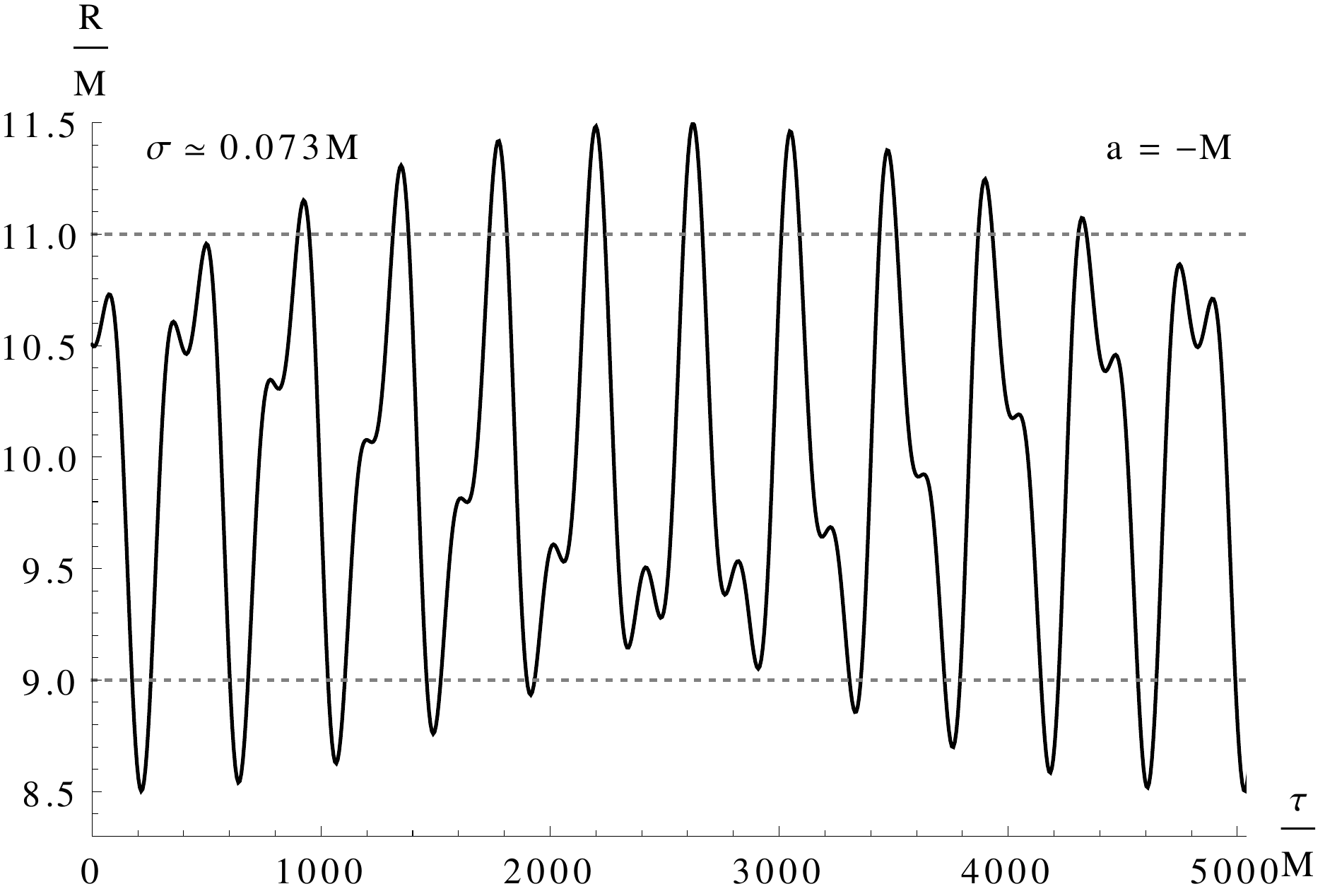}}
\vs{1}

{\footnotesize Fig.\ 1 A particle with prograde spin $\sg\simeq0.073M$ deviating radially from the circular orbit $R=10M$ as a function of  proper time in Kerr field with $a=-M$.}
\ec

While the particle is orbiting closer to the horizons in a co-rotating orbit with $a=0.999M$ in around the circular orbit of radius $R=1.9M$ and with retrograde spin $\sg\simeq-0.052M$, it is tend to go rapid deviations in the radial directions as shown in Fig.\ 2. 
\bc
\scalebox{0.58}{\includegraphics{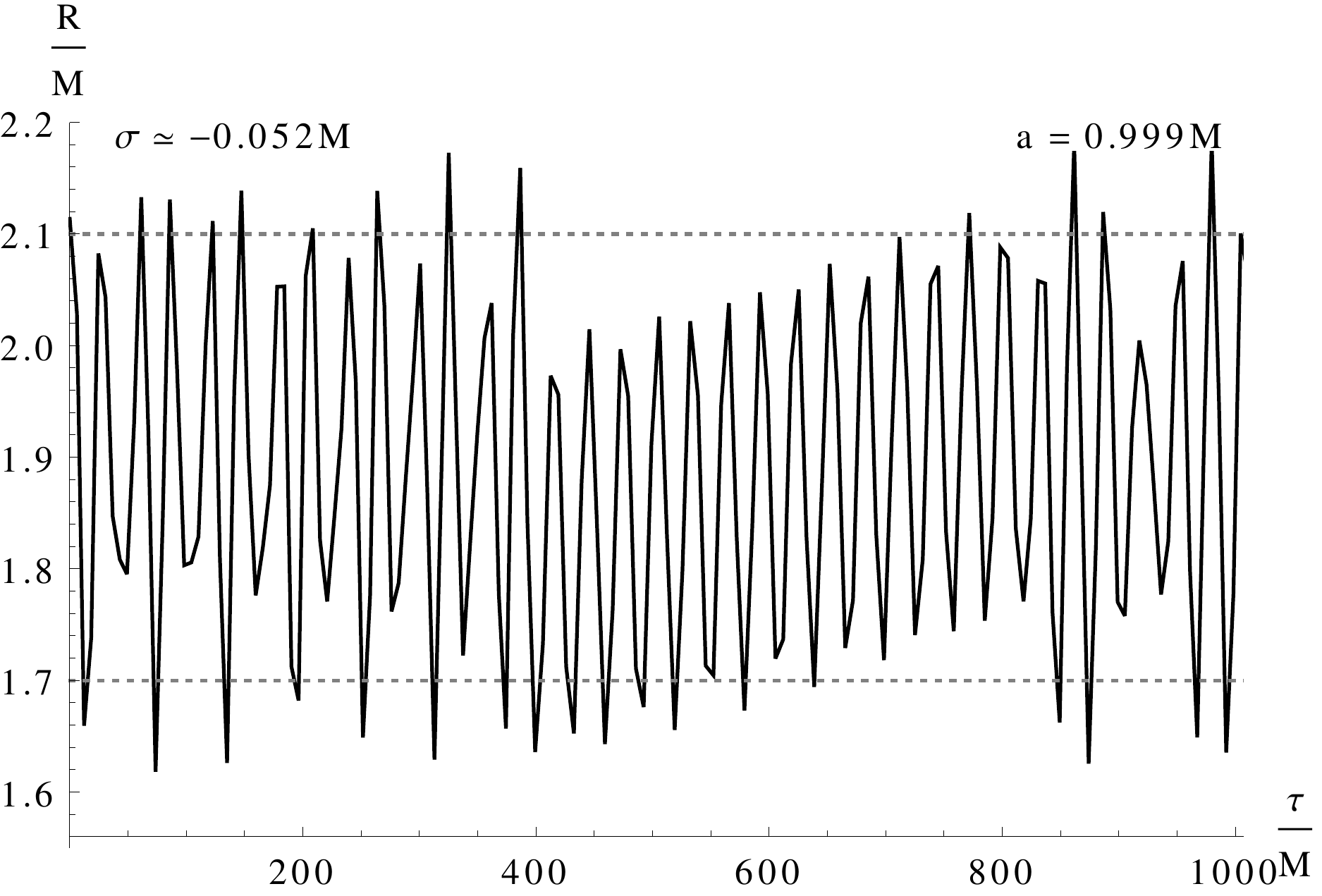}}
\vs{1}

{\footnotesize Fig.\ 2 A particle with retrograde spin $\sg\simeq-0.052M$ deviating radially from the circular orbit $R=1.9M$ as a function of proper time in Kerr field with $a=0.999M$.}
\ec
In both the cases the deviation parameters are chosen as $n^{r}_{+}=0.1R$, $n^{r}_{-}=0.05R$, and $\tau_{+}-\tau_{-} = 100M$. Hence, the particles with spin not only follow the orbital precession but also vary in the radial direction as dictated by the direction and magnitude of the spins of compact binaries.

\section{The ISCO \label{4pisco}}
The minimal radius around the black hole in which the particle can have a bound circular orbit is called the Innermost Stable Circular Orbit (ISCO). For a test body orbiting the non-rotating black hole, it is located at 6M. If either the test body or the black hole possesses angular momentum, then due to spin-orbit coupling, the location of the ISCO varies proportionally to its spin. Thus, for spinning particles in Kerr spacetime, the ISCO depends on the interaction between spin-orbit and spin-spin couplings in the plane.

By analyzing the stability criterion of the deviation solutions from the circular orbits (\ref{nc4}), we predict the ISCO for different values of the Kerr parameter $a$ and the spin angular momentum of the particle $\sg$. The particle will have a bound and periodic orbit only if the frequency $\omega_\pm$ is real. For imaginary frequencies, the circular orbits run into exponential behavior and become unstable \ct{Suzuki:1997by,Blanchet:2002mb}. Thus, from the frequency relation,
\be
\omega^{2}_\pm = \frac{1}{2}\lh A \pm \sqrt{A^2 - 4B}  \rh,
\label{isco1}
\ee
we obtained the conditions:
\be
A^2 - 4B \geq 0,  \hs{1}  A^2 \geq 0, \hs{2}  \textrm{and}  \hs{2} A \geq 0,  \hs{1} B \geq 0,
\label{isco2}
\ee
for $\omega^{2}_\pm$ and $\omega_\pm$ to be real and positive. The intersection of these conditions in the $R$-$\ell^2$ plane separates the stable and unstable orbits. 

The circular orbits in the Kerr plane are determined by the parameters  $\Omega_{\vf}$, $R$, and $\sg$, which fix $\eta$ and $\ve$. Using the conservation laws and the fact that $\Omega_{\vf} = u^{\vf}/u^t$, it can also be expressed in terms of orbital angular momentum, $\ell$, $R$, and $\sg$. Thus, for allowed orbits, fixing $R$ and varying $\ell$ will differ in the value of
\be 
\ba{lll}
 \dsp{ \frac{\sg}{M}}& = & \dsp{\frac{\Sg^{r \vf}}{m M} \lhs \frac{a M (R+M)u^t + (R^3 (R-2M) - a^2 M (2R+M) )u^{\vf} }{a M u^t + (R^2 (R-2M) - a^2 M) u^{\vf} }   \rhs}.\\
 \label{isco3}
\ea
\ee
Where $\ell$, $R$ and $\sg$ are all measured in $M$ units, and $\sg/M$ is calculated by combining equations (\ref{3.3}) and (\ref{3.13}). The intersection of inequalities (\ref{isco2}) for allowed orbits is then investigated in a dimensionless plot as a function of radial coordinate $R$ and angular momentum
\be 
\ba{lll}
 \dsp{\ell}= \dsp{-\frac{2 a M}{R} u^{t} + \lh R^2 + a^2 \lh1+\frac{2M}{R}\rh  \rh u^{\vf}} =  \dsp{ \eta - \sg }.
 \label{isco4}
\ea
\ee
The shaded region denotes the stable circular orbits. Furthermore, the $R$-$\ell^2$ plane contains curves $\ag$, $\beta$, $\gamma$ and $\delta$ that represent isospin lines for spins $\sg = 0$, retrograde spins  $\sg= - 0.55M$, and prograde spins $\sg= 0.55M$ and $0.7M$, respectively. While each of these constant spin lines passes through $B=0$ or $A^{2}-4B=0$, the corresponding radius $R/M$ implies the radius of the ISCO.
\bc
\scalebox{0.48}{\includegraphics{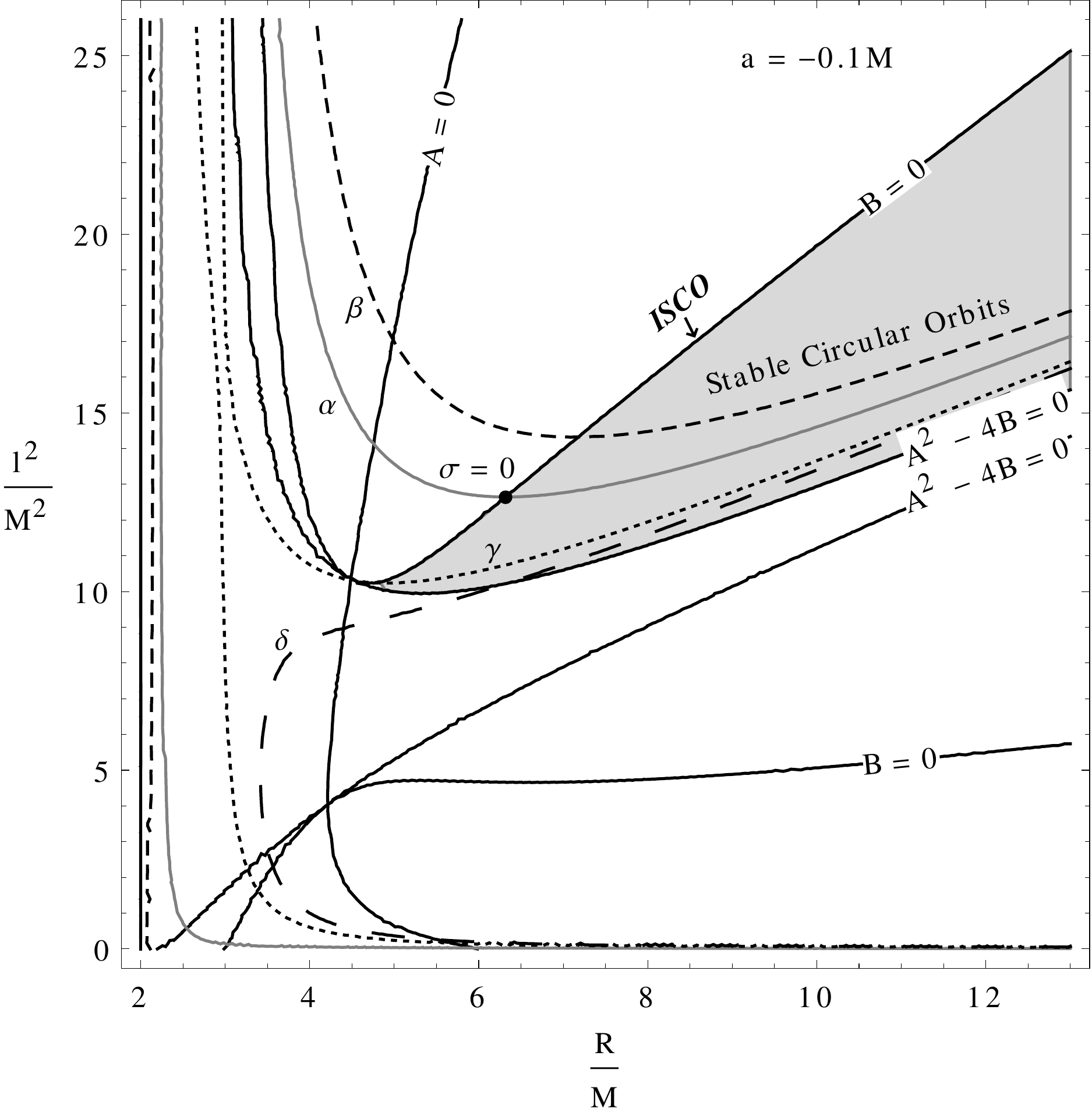}}
\vs{1}

{\footnotesize Fig.\ 3:  Intersection of stability conditions defining plane circular orbits for Kerr parameter $a=-0.1M$ is shown in the $R$-$\ell^2$ plane. The curves labelled  $\ag$, $\beta$, $\gamma$ and $\delta$ indicate orbits of constant spins $\sg/M$  $=$ $0$, $-0.55$ (retrograde), $0.55$ and $0.7$  (progrades), respectively.}
\ec

\bc
\scalebox{0.48}{\includegraphics{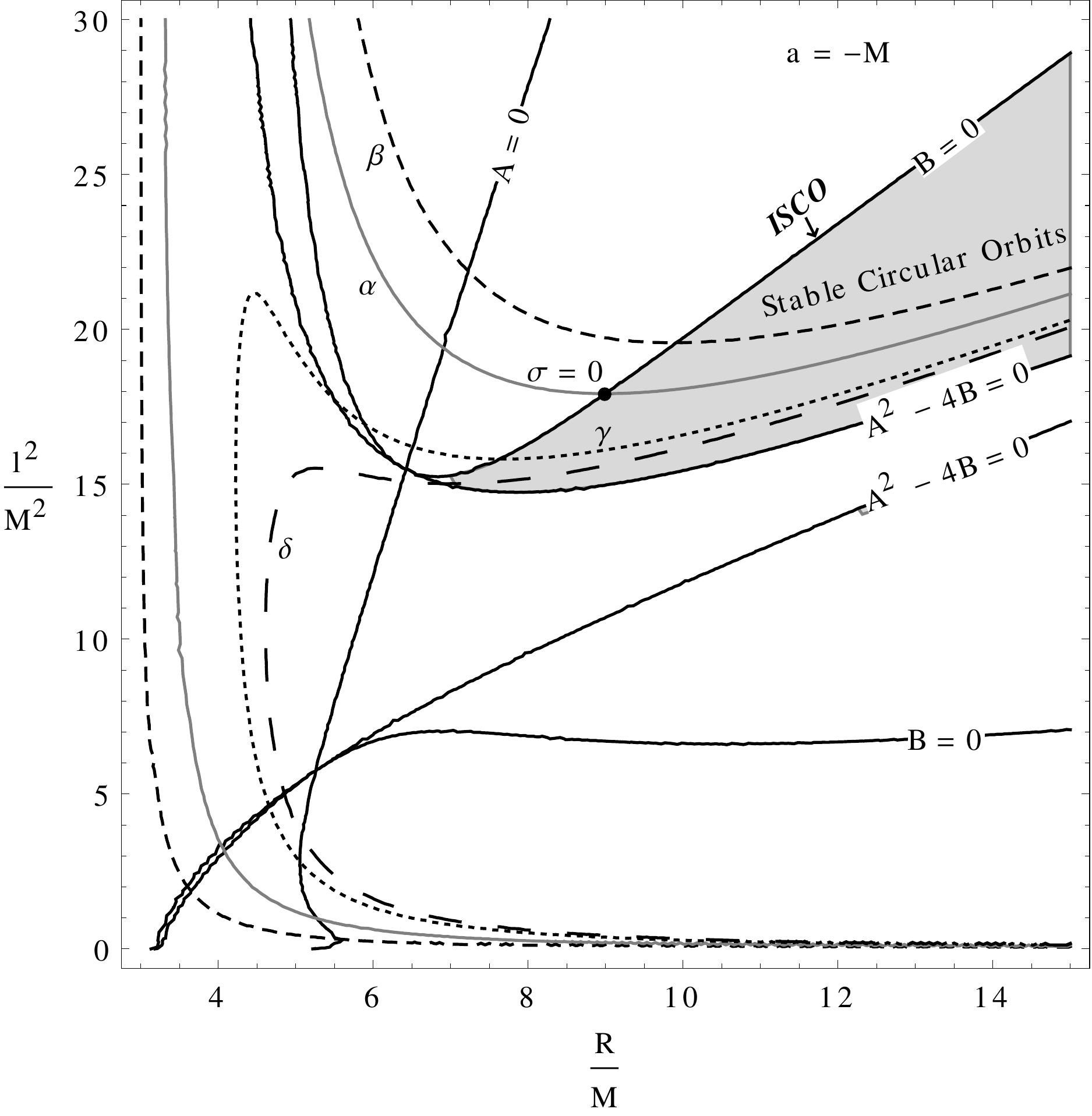}}
\vs{1}

{\footnotesize Fig.\ 4:  Stable circular orbits for Kerr parameter $a=-M$ is given as a function of $R/M$ and $\ell^2/M^2$.}
\ec
 As shown in Fig.\ 3, the vanishing spin $\sg=0$ of a counter-rotating black hole with a Kerr parameter of $a=-0.1M$ is at $R=6.32M$, a known ISCO value for test mass with angular momentum  $\ell^2 = 12.64 M^2$. In the case of retrograde spin, it is at higher values of $R$. For a prograde spin, it is at lower $R$ values, and the smallest ISCO is at $R\simeq4.75M$, where the angular momentum is at its lowest: $\ell^2\simeq10.24 M^2$. The lines cross the condition $A^{2}-4B=0$ when the spin $\sg$ is greater than $0.55M$. The $\delta$ curve shown for $\sg= 0.7M$ is an example. 

In Fig.\ 4. we show that, for $a=-M$, the spin $\sg=0$ line exactly coincides with the ISCO at well-known radius $R=9M$ and corresponding angular momentum $\ell^2 = 17.93 M^2$. Similar to the previous case, the $R$ value increases and decreases for retrograde and prograde spins, respectively. For the prograde spin value of  $\sg=0.64M$, the smallest ISCO exists at $R\simeq7.2M$ with the least angular momentum of $\ell^2\simeq15.39 M^2$. The $\delta$ curve, with $\sg=0.7M$, is closer to the minimum and crosses $A^{2}-4B=0$. 

From these analyses, we obtained the radius of the ISCO as a function of the spin parameter $\sg/M$ in the domain of physical interest: $-0.55<\sg/M<0.55$. The above cases, $a=-0.1M$ and $a=-M$, are illustrated in Fig.\ 5. by curves $c$ and $l$, respectively. Similarly, we predict for eleven different Kerr parameter values ranging from $a = 0$ to $1$ in $-0.1M$ intervals. The curve $b$ represents Schwarzschild ($a=0$)\ct{dAmbrosi:2015wqz}. In all these cases, the constant spin line intersects with the condition $B=0$. It is found that, in the case of counter-rotation, for any given $a$ the radius of the ISCO decreases for prograde spin and increases for retrograde spin as the magnitude of the spin parameter $\sg/M$ increases.

\bc
\scalebox{0.55}{\includegraphics{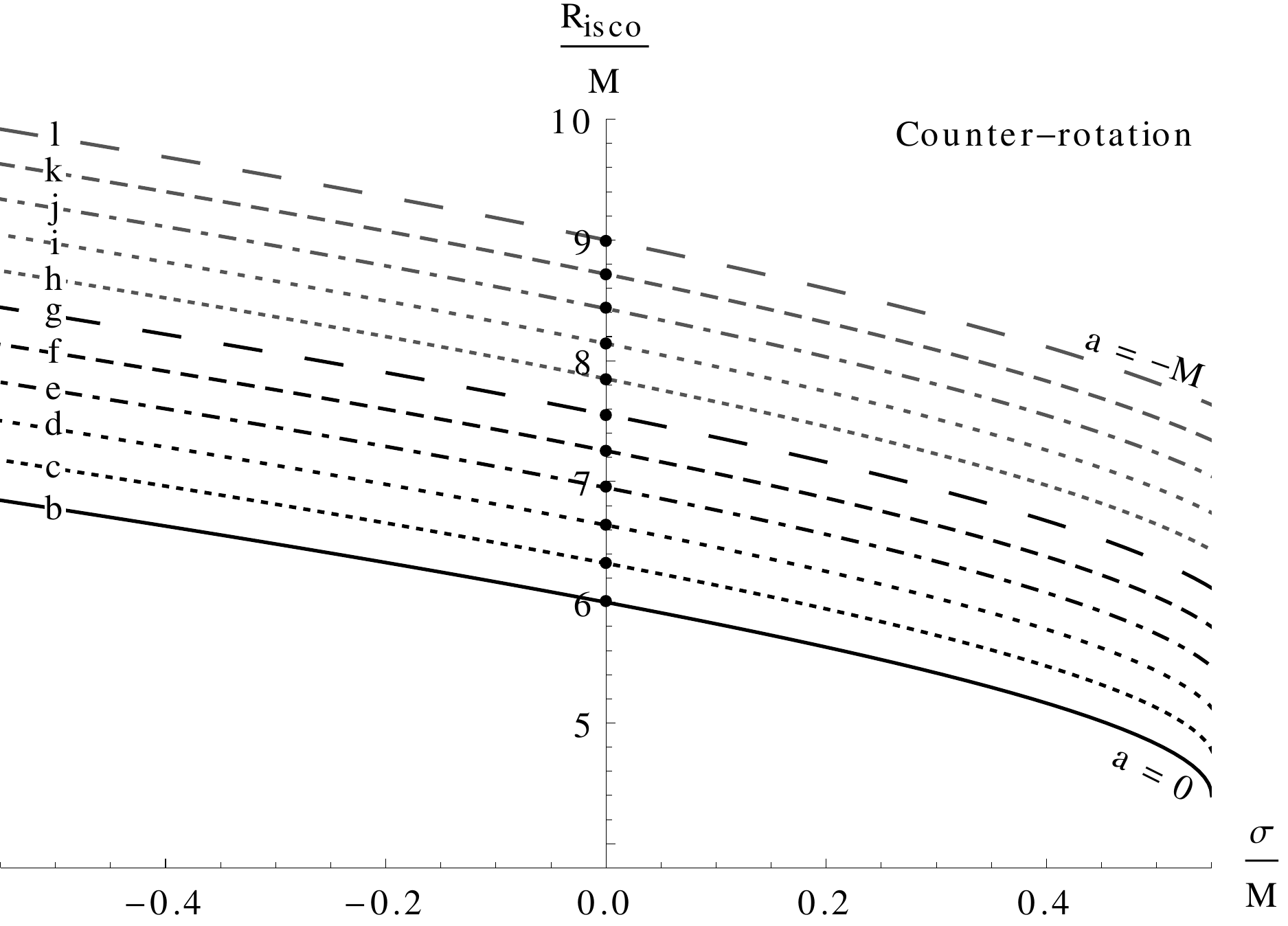}}
\vs{1}
 
{\footnotesize Fig.\ 5: Counter-rotating Kerr black hole. Radius of the ISCO: $R_{isco}$ as a function of spin $\sg/M$ for different values of Kerr parameters. Curves labelled $b (a=0)$,  $c (a=-0.1M)$,  $d (a=-0.2M)$,  $e (a=-0.3M)$,  $f (a=-0.4M)$,  $g (a=-0.5M)$,    $h (a=-0.6M)$,  $i (a=-0.7M)$,  $j (a=-0.8M)$,  $k (a=-0.9M)$ and $l (a=-M)$, representing counter-rotating orbits.}
\ec
\np
A similar analysis can be carried out for co-rotating orbits. We establish solutions to inequalities in the $R$-$\ell^2$ plane for the cases $a=0.1M $ and $0.5M $. It is shown in Fig.\ 6. for a black hole with $a =0.1M$ the vanishing spin $\sg=0$ cross at $B=0$ coincides with the known value of radius $R=5.67M$ and angular momentum $\ell^2=11.34 M^2$. $R$ increases for retrograde and decreases for prograde spins. The smallest ISCO in this case is $R\simeq 4.2 M$, with a minimum angular momentum $\ell^2 \simeq 8.81M^2$ for spin $\sg=0.55M$. Lines with spin $\sg>0.55M$ pass through $A^2 -4B=0$ as indicated with $\delta$.
\bc
\scalebox{0.49}{\includegraphics{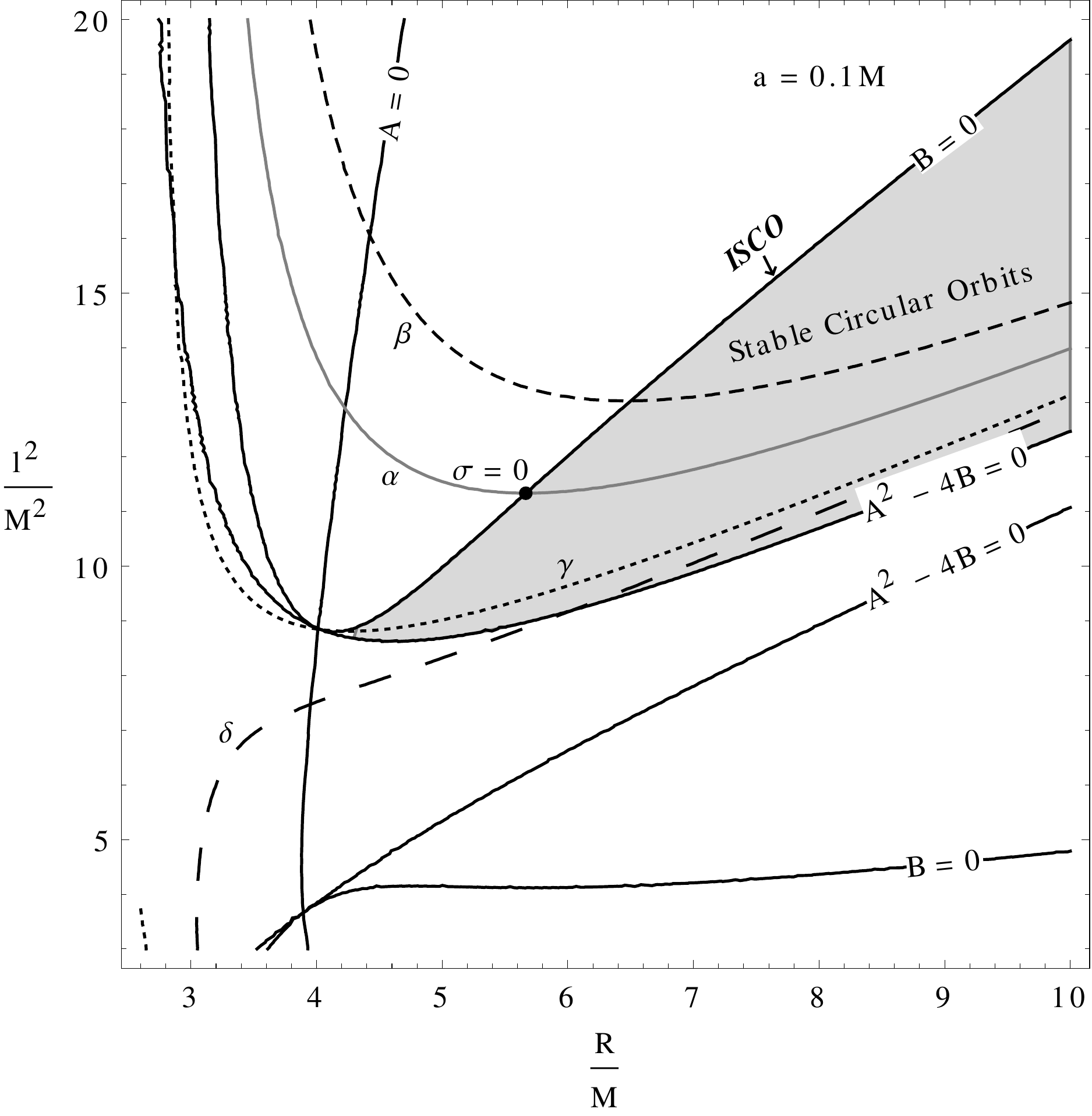}}
\vs{1}

{\footnotesize Fig.\ 6:  Stable circular orbits for Kerr parameter $a=0.1M$ is given as a function of $R/M$ and $\ell^2/M^2$.}
\ec
For $a=0.5M$, Fig.\ 7. shows that the area corresponding to stable circular orbit increases. All of the constant spin curves cross the line $B=0$ indicating the $R$ value of ISCO. When spin $\sg=0$, the radius $R=4.23M$ and angular momentum $\ell^2=8.43 M^2$ coincides with the known values. The $R$ value decreases for prograde spin and increases for $\sg > 0.7M$. The orbit corresponding to the least angular momentum exists at higher prograde spin. In the case of retrograde spin, $R$ increases with an increase in the magnitude of spin.

Figure \ 8. shows a summary of the $R$-$\ell^2$ analysis for the various co-rotating cases where the ISCO radius is proportional to the spin parameter. The curves $b$, $m$, $n$, $o$, $p$, $q$, $r$, and $s$ correspond to Kerr parameter values $a$ $=$ $0$ to $0.7$ in $0.1M$ intervals. We observe that the radius of the ISCO decreases and increases for prograde and retrograde spin, respectively, for an increase in the magnitude of spin $\sg/M$.

Throughout the analysis, the  $R_{isco}/M$ values for vanishing spin, $\sg/M=0$, exactly match those for test bodies  \ct{Bardeen:1972fi}. Even though we have included higher values of the spin parameter $\sg/M$, the ISCO corresponding to such high values will be unreliable because the test particle limit breaks down and we need to consider the masses of two objects comparable. Thus, we only present the results in the domain of physical interest ($-0.55<\sg/M<0.55$).
\bc
\scalebox{0.49}{\includegraphics{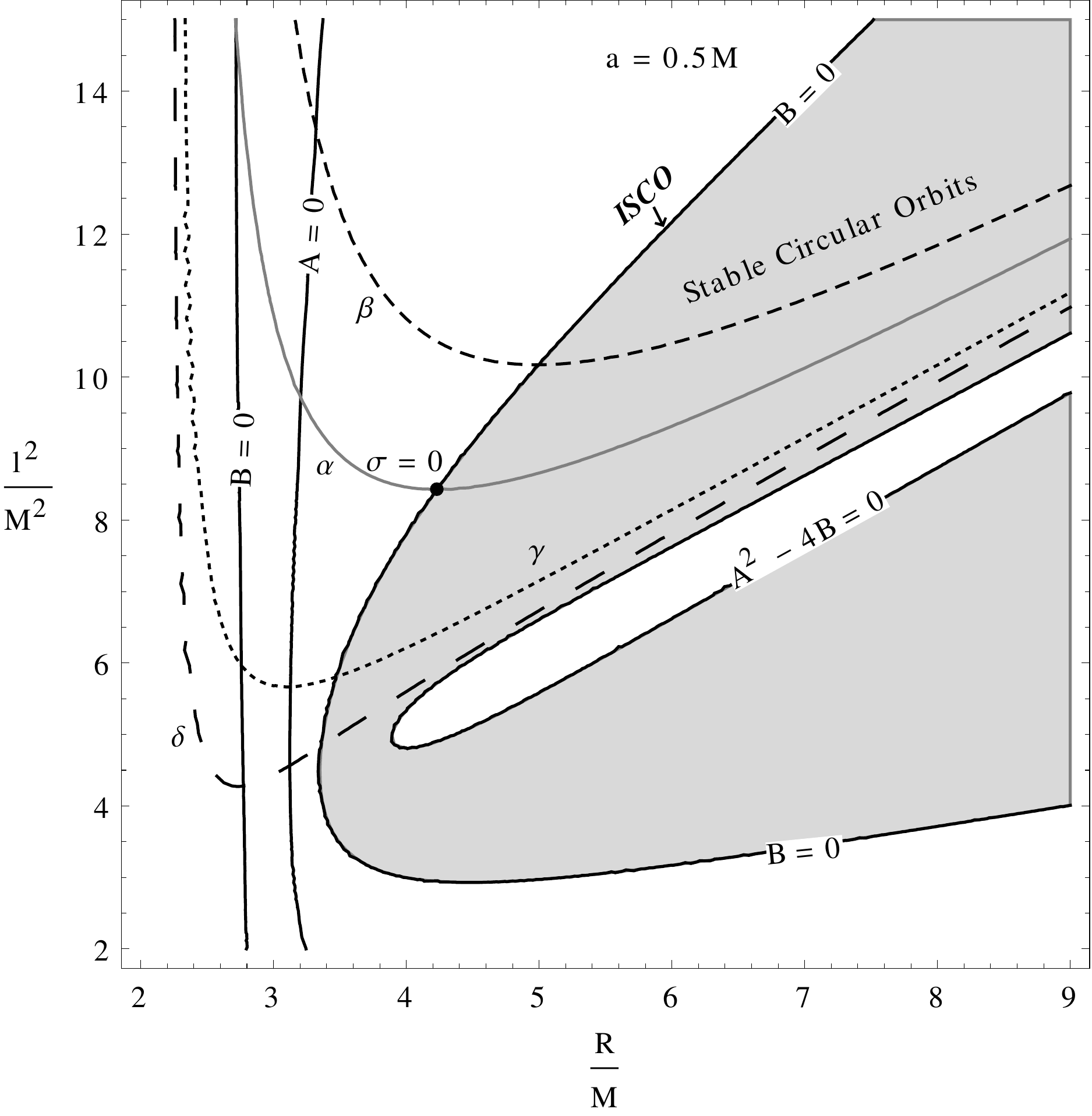}}
\vs{1}

{\footnotesize Fig.\ 7:  Stable circular orbits for Kerr parameter $a=0.5M$ is shown in the $R$-$\ell^2$ plane.}
\ec

\bc
\scalebox{0.55}{\includegraphics{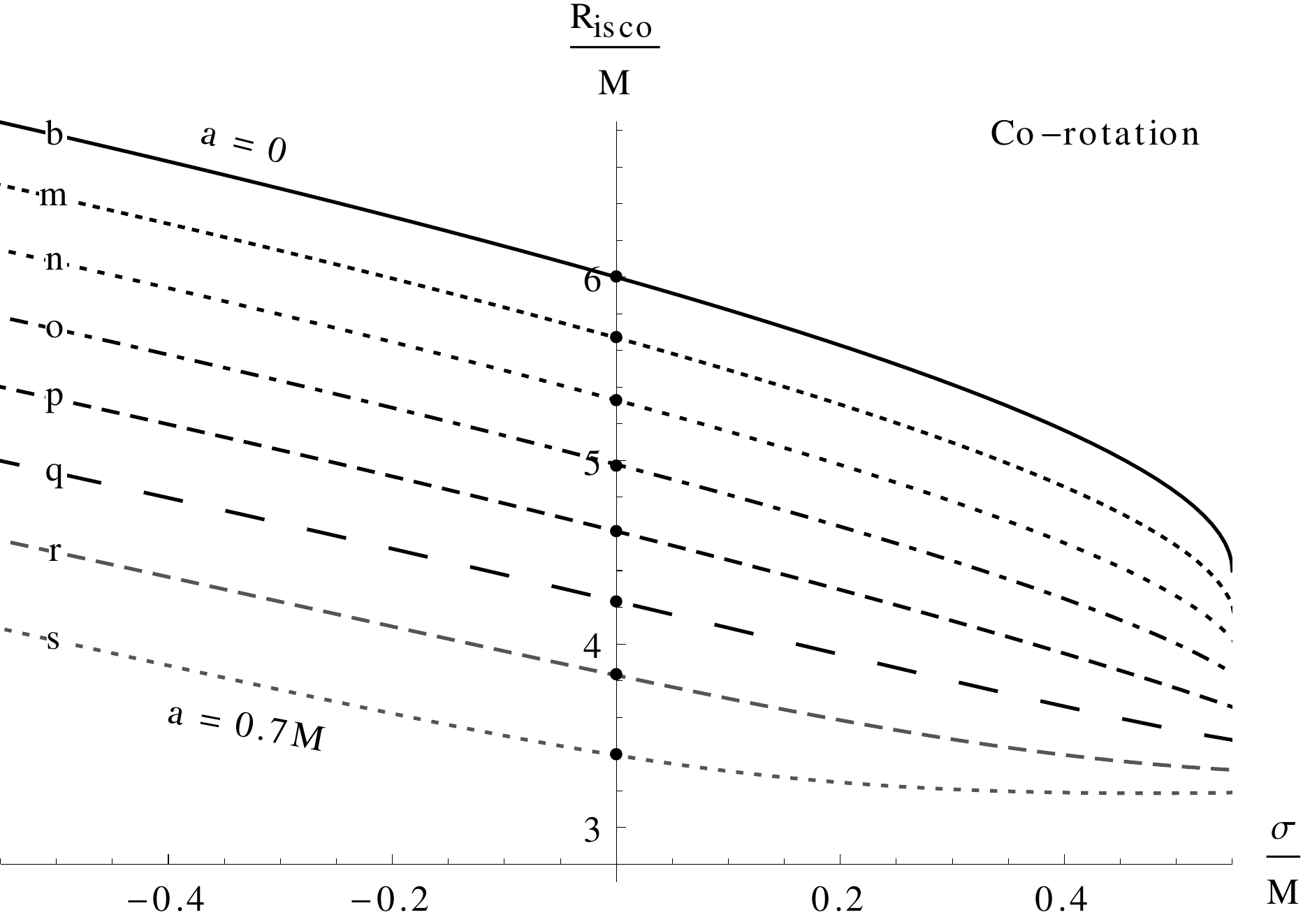}}
\vs{1}

{\footnotesize Fig.\ 8: Co-rotating Kerr black hole. Radius of the ISCO: $R_{isco}$ as a function of spin $\sg/M$ for different values of Kerr parameters. Curves labelled $b (a=0)$, $m (a=0.1M)$, $n (a=0.2M)$, $o (a=0.3M)$, $p (a=0.4M)$, $q (a=0.5M)$, $r (a=0.6M)$, and $s (a=0.7M)$, representing co-rotating orbits.}
\ec
By comparing our results with the existing Mathisson-Papapetrou literature \ct{Suzuki:1997by,Hackmann:2014tga} we found that it agrees qualitatively in the co-rotating region. Whereas it differs in the counter-rotating region due to the fact that we follow the world-line where the spin tensor is covariantly constant and the examples given in the  work \ct{Hackmann:2014tga} follows a specific center of mass.

\section{Precessional frequency of the orbital plane\\in Schwarzschild spacetime\label{}}
In the equatorial configuration, the symmetries of spacetime (Kerr or Schwarzschild) require all the $\theta$ components of the spin tensor to be zero. Thus, the direction of the spin and orbital angular momentum need to be aligned, and therefore the particles' motion is confined to within a plane. However, in the generic case, the coupling of the particle's spin with the background geometry enforces the precession of the orbit \ct{Bini:2014poa,Ruangsri:2015cvg}. The non-availability of the Carter constant of motion \ct{Rudiger:1983} for the complete spin of the particle makes it difficult to develop the generic orbits in the Kerr spacetime. Hence, we specialize in developing completely relativistic non-planar eccentiric orbits around the Schwarzschild spacetime without truncating the particle's spin. 

The constants of motion make it easy to solve the deviation equations. The conserved quantities emerging out of spherical symmetry as implied by the Noether's theorem:
\be 
\ba{lll}
\dsp{\ve} =  \dsp { \lh 1 - \frac{2M}{r} \rh u^{t}  - \frac{M}{m r^2} \Sg^{tr},}  \hs{3}   \eta   =   r^2 u^{\vf}  + \frac{r}{m} \Sg^{r \vf}, \\
& & \\
\Sg^{r\thg}  =  -mru^{\thg},    \hs{3}    \dsp{\Sg^{\thg\vf}= \frac{J}{r^2}\ctg{\thg}}. 
 \label{4.1}
\ea
\ee
And additionally the universal constant $D$ from equation (\ref{2.7}) turns out to be an equation for the spin component $\Sg^{t\thg}$:
\be
\ba{ll}
\dsp{\lh \eta -r^2\, u^{\vf}\rh \Sg^{t\thg}  +r^2 u^{\thg}\,\Sg^{t\vf}} = \dsp{-\frac{D}{m r}\csc^{2}{\thg}  + \frac{m\eta}{M} \lh (r-2M)\,u^{t} - r \ve \rh\, \ctg{\thg}}.
 \ea
  \label{4.2}
\ee
In this setting, the system of deviation equations deduces two sets of coupled first order differential equations. The first set consists of orbital and spin parameters, $(\del t, \del r, \del\vf)$, and $(\del\Sg^{t\vf}, \del\Sg^{tr}, \del\Sg^{r\vf})$, describing planar deviations. The second set contains parameters $\del u^{\thg}$ and $(\del\Sg^{t\thg}, \del\Sg^{r\thg}, \del\Sg^{\thg\vf})$ for orbital and spin degrees of freedom, respectively, characterizing non-planar orbits. Starting from the circular referance orbit, these two sets can be solved independently. Since we have already solved the planar deviations in Schwarzschild \ct{dAmbrosi:2015wqz}, and generalized the same in the section \ref{npco} for Kerr, here in the following we solve the non-planar deviations \ct{Saravanan:2016zkm}.

By evaluating the coefficients at the referance orbit with the necessary conditions (\ref{3.01}) and  (\ref{3.16}), the spin deviation equations are expressed as
\be
\ba{lll}
\dsp{\frac{d{\del \Sg^{t\thg}}}{d\tau}}  &=&  \dsp{-\frac{M u^t}{R(R-2M)}\, \del\Sg^{r\thg} - \frac{\Sg^{tr}}{R}\, \del u^{\thg}},\\
&&\\
\dsp{\frac{d{\del \Sg^{r\thg}}}{d\tau}} &=& \dsp{-\frac{M(R-2M)u^{t}}{R^3}\,  \del \Sg^{t\thg}  - (R-2M) u^{\vf} \, \del \Sg^{\thg\vf} - u^{\vf}\Sg^{r\vf} \del\thg}, \\
&& \\
\dsp{\frac{d{\del \Sg^{\thg\vf}}}{d\tau}}  &=&  \dsp{-\frac{\Sg^{r\vf}}{R}\, \del u^{\thg} + \frac{u^{\vf}}{R}\,  \del \Sg^{r\thg} }.\\
\ea
\label{4.3}
\ee
It is worth noting that these equations  (\ref{4.3}) are identical to the conservation laws (\ref{4.1}) and  (\ref{4.2}). As a result, by calculating the remaining orbital deviation for $\del\thg$,
\be
\ba{lll}
   \dsp{\frac{d^2\del \theta}{d \tau^2}} +   \dsp{  \Bigg[ - \frac{2M\eta}{R^3}u^{\vf}  + \lh 1 + \frac{(R-2M)}{R (\eta - R^2 u^{\vf})} \eta    \rh u^{\vf\,2}+ \frac{\eta (R-2M)  \lh 1 - \ve u^{t} \rh}{R^3 (\eta - R^2 u^{\vf})} \Bigg] \del \thg}\\ 
& & \\
\hs{21} =  \dsp{ -  \lhs \frac{M (R-2M) u^{t}}{m^2 R^5 (\eta - R^2 u^{\vf})} \rhs \del D},\\
&& \\
\ea
\label{4.4}
\ee
 we compute the non-planar spin deviations. The solution of the harmonic oscillator equation (\ref{4.4}) with the constant driving force proportional to $\del D$ is 
\be
\ba{l}
 \dsp{\del \thg (\tau)}  =  \dsp{ \thg_{(\tau=0)} \cos{\Og_{p} \tau}  + \frac{u^{\thg}_{(\tau=0)}}{\Og_{p}}  \sin{\Og_{p} \tau    -  \frac{M (R-2M) u^{t} }{m^2 R^5\,\Og^{2}_{p} (\eta - R^2 u^{\vf})}\, \del D}},
\ea
\label{4.5}
\ee
wherein,  $\thg_{(\tau=0)}$ and $u^{\thg}_{(\tau=0)}$ are the position and angular velocity of the particle at the initial (proper) time $\tau=0$, and $\Og_{p}$ denotes the frequency of the precessing orbital plane,
\be
\ba{ll}
\dsp{\Og_{p}}=\dsp{\sqrt{- \frac{2M\eta}{R^3}u^{\vf}  + \lhs 1 + \frac{\eta (R-2M)}{R (\eta - R^2 u^{\vf})}     \rhs u^{\vf\,2}+ \frac{\eta (R-2M)  \lh 1 - \ve u^{t} \rh}{R^3 (\eta - R^2 u^{\vf})}  }}.  \\
& \\
\ea
\label{4.6}
\ee
Thus, we have derived the precessional frequency in terms of the circular orbit constants $(R, u^{t}, u^{\vf})$ and conserved quantities $(\ve,\eta)$. We found that for vanishing spin $\sg =0$, $\Og_p$ is the same as the test mass's azhimuthal frequency:
\be
\Og_{p} = \Og_{\vf}  = \sqrt{\frac{M}{R^3}}
\ee
which implies, the plane of the orbit is no longer precessing. 

Since we obtain the angular variation $\del\thg$, it is  straight forward to compute the spherical deviations by using the conserved quantities (\ref{4.1}) and  (\ref{4.2}). The complete equations will be of the form:
\be
\ba{l}
\thg(\tau) = u^{\thg} \tau + \del\thg (\tau), \\ 
 \\
 \Sg^{r\thg}(\tau) = \Sg^{r\thg}_{0} +\del \Sg^{r\thg}(\tau) ,  \\
 \\
 \Sg^{\thg\vf}(\tau) = \Sg^{\thg\vf}_{0} + \del \Sg^{\thg\vf}(\tau),  \\
 \\
\Sg^{t\thg}(\tau) = \Sg^{t\thg}_{0} + \del \Sg^{t\thg}(\tau).
\ea
\label{}    
\ee
Thus, we write the complete first order world line approximations for non-planar orbits after evaluating the velocity $u^{\thg}$ and spin components $\Sg^{r\thg}_{0}$, $ \Sg^{\thg\vf}_{0}$, and $ \Sg^{t\thg}_{0}$ at circular orbits: 
\be
\ba{lll}
\thg(\tau)  & = &  \dsp{ \thg_{(\tau=0)} \cos{\Og_{p} \tau}  + \frac{u^{\thg}_{(\tau=0)}}{\Og_{p}}  \sin{\Og_{p} \tau    -  \frac{M (R-2M) u^{t} }{m^2 R^5\,\Og^{2}_{p} (\eta - R^2 u^{\vf})}\, \del D}}, \\
 && \\
 \Sg^{r\thg}(\tau) &=& \dsp{m R\,  \Og_{p} \lhs  \thg_{(\tau=0)} \sin{\Og_{p} \tau}   - \frac{u^{\thg}_{(\tau=0)}}{ \Og_{p}}  \cos{\Og_{p} \tau}   \rhs}, \\
 && \\
 \Sg^{\thg\vf}(\tau)&=&\dsp{-\frac{m \eta}{R^2}  \lhs  \thg_{(\tau=0)} \cos{\Og_{p} \tau}  + \frac{u^{\thg}_{(\tau=0)}}{\Og_{p}}  \sin{\Og_{p} \tau    -  \frac{M (R-2M) u^{t} }{m^2 R^5\,\Og^{2}_{p} (\eta - R^2 u^{\vf})}\, \del D} \rhs},  \\
 && \\
\Sg^{t\thg}(\tau)  & = &     \dsp{-  \frac{m \eta \lh (R-2M) u^{t} - R \ve \rh}{M (\eta - R^2 u^{\vf})} \lhs  \thg_{(\tau=0)} \cos{\Og_{p} \tau}  + \frac{u^{\thg}_{(\tau=0)}}{\Og_{p}}  \sin{\Og_{p} \tau}  \rhs}\\ 
& & \\
&   &   \dsp{ -\frac{ 1}{m R\, (\eta - R^2 u^{\vf})} \lhs 1 - \frac{\eta (R-2M) \lh 1 - \ve u^{t} + R^2 u^{\vf\,2}  \rh}{R^3 \Og^{2}_{p}  (\eta - R^2 u^{\vf})}  \rhs \del D}.\\
\ea
\label{}    
\ee

\bc
\scalebox{0.58}{\includegraphics{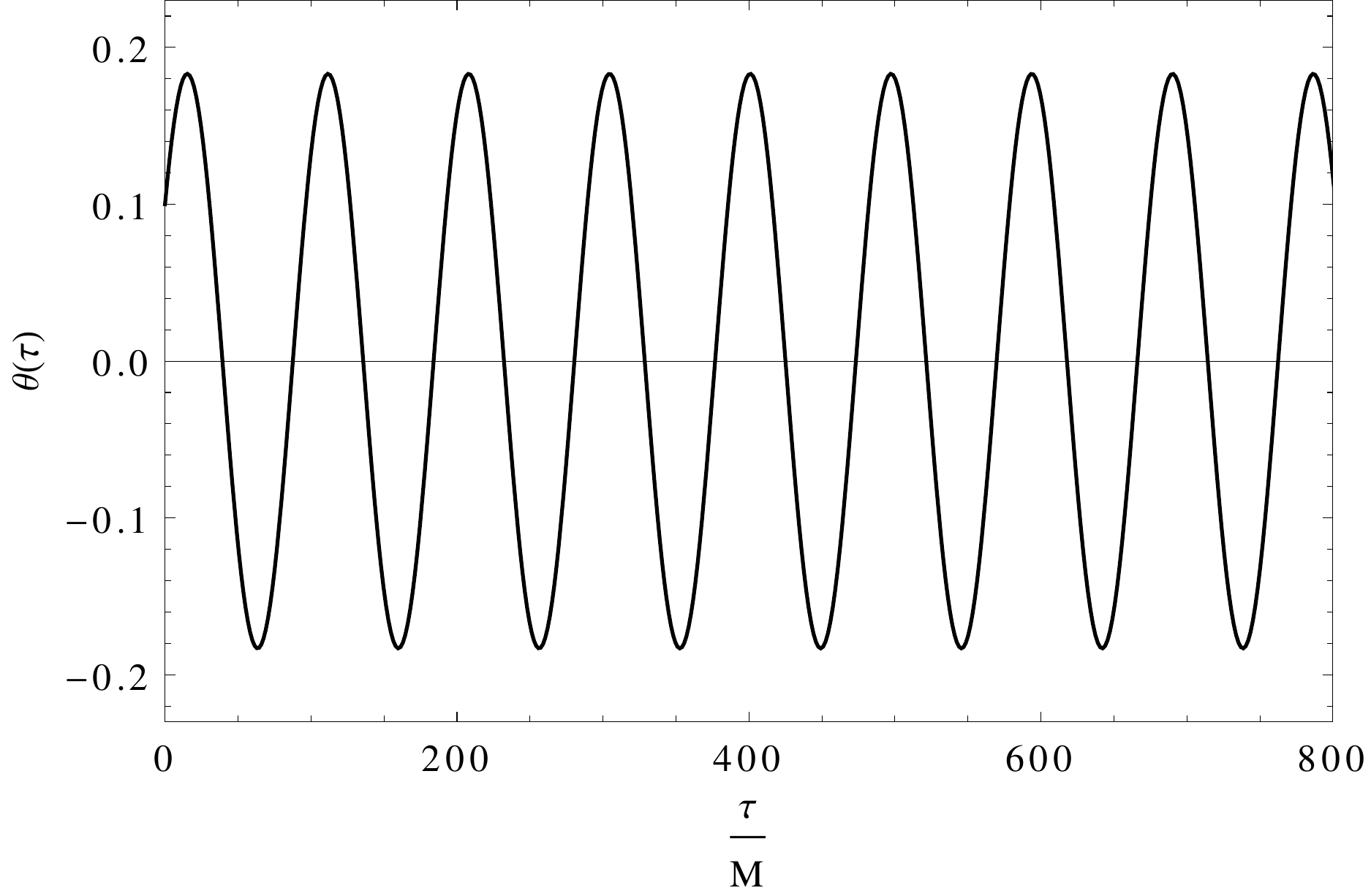}}

\vs{1.11}
{\footnotesize Fig.\ 9: The dimensionless plot  $\thg (\tau)$ versus $\tau/M$ in Schwarzschild spacetime.}

\ec
We conclude this section by plotting (Fig.\ 9) the angular variation $\thg (\tau)$ as a function of the proper time $\frac{\tau}{M}$ for a special case $\del D =0$. Here, we have choosen the orbital radius $R=8M$, and orbital and spin angular momentum magnitudes, $\ell=3.5M$ and $\sg=0.1M$, respectively, with the initial conditions $\thg_{(\tau=0)} = 0.1$ and $u^{\thg}_{\tau=0}=0.01$. Thus, the precessional orbital motion of a spinning particle due to spin-orbit coupling in Schwarzschild spacetime is presented in a completely relativistic framework.

\section{Conclusion and Discussion \label{}}
Various effects of spin have been observed by applying the Covariant hamiltonian formalism to the case of spinning bodies in the test-particle limit ($m<<M$) in the Kerr and Schwarzschild spacetimes. The \emph{exact} solution (\ref{3.19}) for frequency $\Og_{\vf}$ for a particle orbiting the Kerr spacetime in circular trajectories with (\emph{full}) spin $\sg$ is given as a function of radius $R$, alignment and magnitude of spins $\sg$ and $a$. Spinning particles in the planar orbits is tend to align either parallel or anti-parallel w.r.t orbital angular momentum. A large class of plane noncircular bound orbits (\ref{nc4}) has been developed around the exact circular solution by using the world line perturbation theory. It is found that, in addition to the angular shift, the periastron radius varies proportionally to their spins ($\sg, a$), as demonstrated for a particle with a prograde spin in a counter-rotating orbit ($a=-M$) and a particle with a retrograde spin in a co-rotating orbit ($a=0.999M$). The radius of the ISCO as a function of spin $\sg$ has been determined using the stability analysis method for many co-rotating and counter-rotating examples in the region of primary physical interest $-0.55<\sg/M<0.55$ in the Kerr plane (as the spin of the particle $\sigma$ is normalized with the mass of the black hole $M$, the quantity $\sigma/M$ is very small). For generic spin alignments, the perturbative construction has also been extended to nonplanar bound orbits in Schwarzschild geometry, and the precessional frequency of the orbital plane is derived analytically (\ref{4.6}) in a completely relativistic framework.

All these results will have astrophysical implications \ct{Li:2019peo,Miller:2014aaa,Reynolds:2019uxi,Fragione:2020khu}. An extension of the planar analysis of the ISCO for $a > 0.7 M$ is left for detailed future work. It is essential to develop this framework to understand the nonplanar orbits and the precession of the orbital plane in Kerr geometry. Since this approach accounts for a large class of hamiltonians, the dynamics due to the gravitational Stern-Gerlach interactions encoded in the nonminimal hamiltonian \ct{dAmbrosi:2015wqz} can be developed for the practical cases. Further, incorporating the back reaction of the test body and determining the self-force \ct{Akcay:2019bvk} and gravitational waves \ct{Koekoek:2011mm,dAmbrosi:2014llh} will be of fundamental interest for the LISA mission.

\vs{4}
\nit
{\bf Acknowledgements} \\
\\
The author is grateful to Riccardo Sturani for several discussions and thought-provoking questions while developing this work. Indebted to Jan-Willem van Holten (JWvH), Giuseppe d'Ambrosi, and Jorinde van de Vis for their collaborations in developing the formalism \ct{d'Ambrosi:2015gsa,dAmbrosi:2015wqz} used here. Discussions and constructive comments from JWvH are also appreciated. N. D. Hari Dass is sincerely acknowledged for inspiring the author on this subject with a decade-old summer school and internship on ``rotations'', in Bangalore. ``This work was supported by a grant from the Simons Foundation (Grant Number 884966, AF)'', and the research is carried out at the Associação Instituto Internacional de Física. It was partly supported by the ``Salahuddin Foundation'', Dubai.

\section*{Appendix A: Symbols and Dimensions\label{appa}}
The list of different quantities and  their dimensions appearing throughout the text are collected here. We work in the geometrized units (c = G = 1).
$$\begin{array}{rr|l}
\mbox{}  & \mbox{Symbol}   & \mbox{Dimension} \\
\hline
\mbox{} &  (\thg, \vf,  u^{t}, u^{r},  \ve, \Sg^{\thg\vf})  & M^0 \\
\mbox{}& (m, r, a, t, \tau, \eta,  \ell,  \sg, \Sg^{r\vf}, \Sg^{r\thg}, \Sg^{t\thg}, \Sg^{t\vf})  & M  \\
\mbox{}  & (u^{\thg}, u^{\vf}, \Og_{\vf}, \Og_{p}) &  M^{-1}    \\
\mbox{} & (\Del^2, \Sg^{tr}) &   M^2  \\
\mbox{}   & (I, D) &  M^4   \\
\end{array}$$
\section*{Appendix B: Coefficients of the circular orbits equation\label{ab}}
The circular orbits of the spinning particles in Kerr spacetime are described by the equation (\ref{3.18}). Here we assemble the explicit expressions for the coefficients $X$, $Y$, and $Z$, which are functions of radius $R$ and spins $(\sg, a)$: 
\be 
\ba{lll}
X  & = &  a \ve \lh R^3+a^2(R+2M) \rh  \bigg[ 3R^2(R-M) + a^2(3R+M) \bigg] \\
 & &  -\eta \bigg[R^5(R-2M) + a^2 R^2(R^2+5M R - 4 M^2) + a^4 M(5R+2M) \bigg] \\
 & & + R \sg \bigg[ R^4(R-3M) + a^2 R (R^2+3M R - 6M^2) + 4a^4 M  \bigg], \\
 & & \\
 Y & = & -\ve \bigg[2R^5 (R-2M) + a^2 R^2 \lh 5R^2 + 7 M R -14M^2 \rh + a^4 (R+4M)(3R+M) \bigg]\\
 & & - a \eta \bigg[ R^2(R-2M)(3R-5M) + a^2(3R^2-9M R -4M^2) \bigg]\\ 
 & & + a R \sg \bigg[ R(3R^2-10 M R + 12 M^2) + a^2 (3R-8M)   \bigg],\\ 
 & & \\
 Z & = & 2 a \ve \bigg[2 M R^2(R-2M) + a^2 M(3R+M) \bigg]\\
 & & + \eta \bigg[ R^2(2R^2 - 7 M R + 6M^2) + a^2 (3R^2 - 4M R -2M^2) \bigg]\\
 & & - R \sg \bigg[R(2R^2 - 7M R + 6M^2) + a^2 (3R-4M) \bigg].\\
 \label{}
\ea
\ee
\section*{Appendix C: Coefficients of the worldline deviations\label{appc}}
The system of equations accounting for noncircular orbits in the equatorial plane of the Kerr metric is given by equation (\ref{mat}). The explicit coefficients of the corresponding differential equations are collected below. These coefficients reduce to the special cases: for Kerr parameter $a=0$, it reduces to spinning particles in Schwarzschild \ct{dAmbrosi:2015wqz}, and for particle spin $\sg=0$, it reduces to the test mass in Kerr \ct{Colistete:2002ka}. 
\be
\ba{lll}
\nonumber
\dsp{e} &=& \dsp{-\frac{M}{R^4} \big\{ au^t - (R^2+a^2) u^{\vf}  \big\}},     \qquad  \dsp{g} = \dsp{-\frac{m}{ \Del^2 R} \big\{ M u^t - a  (R+M) u^{\vf}  \big\}},
\\
&&\\
\dsp{f} &=& \dsp{\frac{1}{R^4} \big\{u^t \lhs 2R^2 (R-2M) + a^2 (3R+M) \rhs  - a u^{\vf} \lhs 3R^2 (R-M) + a^2 (3R+M)  \rhs\big\}}, \\
&&\\
\dsp{h} &=& \dsp{\frac{m}{ \Del^2 M R} \big\{ a M (R+M) u^t + \lhs R^3 (R-2M) - a^2 M (2R+ M) \rhs u^{\vf}  \big\}}, \\
&&\\
\dsp{\bg} &=& \dsp{ - \frac{M }{m R^4} \lhs R^3 + a^2 (R+2M) \rhs u^{\vf}},   \qquad  \dsp{\zeta} =  \dsp{-\frac{M \lh R-2M \rh}{m R^4}u^{t}}, \\
&&\\
\dsp{\ag} &=& \dsp{ \frac{1}{\Del^4 R^3} \big\{ u^t \big[ 2 R^4 (R-M)(R-2M) + a^2 R \lh 5R^3 - 5MR^2 -12 M^2 R + 16M^3 \rh}\\
\vs{-3} \\
&& \dsp{ \hs{4.5} + a^4 \lh 3R^2 + 3MR - 10M^2 \rh \big]} \\
\vs{-2} \\
&&\dsp{\hs{3} + aM u^{\vf} \lhs -R^4 (3R-4M) + 2a^2 M R (R-8M) + a^4 (3R+10M)\rhs} \\
\vs{-2} \\
&&\dsp{\hs{3} - \ve \lhs 2R^5 (R-2M) + a^2 R^2 (5R^2 + MR - 8M^2) +a^4 (3R^2 + 7MR +2M^2)  \rhs} \\
\vs{-2} \\
&&\dsp{\hs{3} + aM\eta \lhs R^3 + a^2 (R+2M) \rhs \big\}}, \\
&& \\
\dsp{\gam} &=& \dsp{ \frac{1}{\Del^4 R^3} \big\{ a u^t \lhs R (R-2M) (3R^2 - 7 M R + 8 M^2) + a^2 (3R^2 - 9M R + 10 M^2)\rhs }\\
\vs{-2} \\
 &&\dsp{ \hs{3} + u^{\vf} \big[ R^4 (R-2M) (2R-5M) + 2a^2 R (R-2M) (R^2 + M R - 4M^2)  } \\
 &&\dsp{  \hs{5.2} + a^4 M (3R-10M) \big] } \\
 \vs{-2} \\
 &&\dsp{ \hs{3} -a \ve  (R-2M) \lhs 3R^2 (R-M) + a^2 (3R+M)  \rhs } \\
 \vs{-2} \\
 && \dsp{\hs{3} + M \eta (R-2M) (R^2 + a^2) \big\} }, \\
 && \\
 \dsp{\kg} &=& \dsp{\frac{1}{R^4 (R-2M)} \big\{-a \Del^2 R (3R-2M) u^{\vf} + a M  (R-2M) \eta  } \\
&&\dsp{\hs{6.6} - \ve  (R-2M)\lhs 2R^2 (R-2M) +a^2 (3R+M) \rhs \big\}}, \\
&&\\
\dsp{\lb} &=& \dsp{\frac{1}{R^4 (R-2M)} \big\{ -aR \Del^2 (3R-2M) u^t   - M \eta (R-2M) (R^2+a^2)  } \\
&& \dsp{\hs{6.6} + 2 u^{\vf} R \lhs R^4 (R-2M) + 4 a^2 R (R-M)^2+ a^4 (3R-2M)  \rhs } \\
\vs{-2} \\
&& \dsp{\hs{6.6} + a \ve (R-2M) \lhs 3R^2 (R-M) + a^2 (3R+M) \rhs \big\} }, 
\ea
\ee

\be
\ba{lll}
\dsp{\nu} & = & \dsp{\frac{m}{\Del^2 M R (R-2M)} \big\{ R (R-M) u^{\vf} \lhs R (R-2M)(R-3M) - 2 a^2 M  \rhs} \\
&&\dsp{\hs{8} -aM (R+M) (R-2M) \ve + M^2 \eta  (R-2M)    \big\}}, \\
&& \\
\dsp{\sg} & = & \dsp{\frac{m}{\Del^2 M R (R-2M)} \big\{R (R-M) u^t \lh R(R-2M) (R-3M) - 2 a^2 M \rh  } \\
\vs{-3.5} \\
&&\dsp{\hs{8} + 4 a M R (R-M)u^{\vf} (R(2R-3M) +a^2) } \\
\vs{-2} \\
&&\dsp{\hs{8} - \ve  (R-2M) \lh R^3 (R-2M) - a^2 M (2R+M) \rh } \\
\vs{-2} \\
&&\dsp{\hs{8}  - a M \eta  (R+M)  (R-2M)   \big\}}, \\
&& \\
\dsp{\mu}&=&\dsp{\frac{1}{R^5 (R-2M)^2} \big\{ -R \lhs 2 R(R-3M)(R-2M)^2 + a^2 (9R^2 - 28M R +24 M^2) \rhs } \\
&&\dsp{\hs{6.6} + a R u^t u^{\vf} \big[ R (3R-4M) (R-2M)^2 + a^2 \lh 9R^2 -20M R +12 M^2  \rh \big] } \\
\vs{-2} \\
&&\dsp{\hs{6.6} + R u^{\vf\,2}\big[ R^4 (R-2M)^2 - 4 a^2 R(R-M) (R^2-3M R+4M^2) } \\
&&\dsp{\hs{9.3}-a^4 (9R^2 - 20 M R +12 M^2)  \big]}\\
\vs{-2} \\
&&\dsp{\hs{6.6} + (R-2M)^2 u^t \lhs -4aM\eta + \ve \lh 2R^2 (R-4M) + a^2 (9R+4M) \rh  \rhs} \\
\vs{-2} \\
&&\dsp{\hs{6.6} + (R-2M)^2 u^{\vf} \big[ 2M\eta (R^2+2a^2)   }\\
&&\dsp{\hs{13.1} - a\ve \lh 3R^2 (R-2M) + a^2 (9R+4M)   \rh \big]}, \\
&& \\
\dsp{\chi}&=&\dsp{\frac{m}{MR^2 \Del^4 (R-2M)^2}\big\{ - aM R^2 \lhs \lh 2R^3 -7M R^2 + 8M^2 R-4M^3 \rh + a^2 M  \rhs } \\
\vs{-2} \\
&&\dsp{\hs{4.3} + R^2 u^t u^{\vf} \big[R^2 (R-2M)^2 \lh R^2 - 4MR + 5M^2 \rh + 2a^4 M^2  }\\
&&\dsp{\hs{9} + a^2 (R-2M) (3R^3 -10 M R^2 + 13 M^2 R - 2M^3)  \big]} \\
\vs{-2} \\
&&\dsp{\hs{4.3} -2 a M R^2 u^{\vf\,2} \big[ M R^2(R-2M)(3R-4M) + a^4 M }\\
&&\dsp{\hs{11} - 2a^2 \lh R^3 - 5 M R^2 + 6M^2 R-M^3 \rh  \big]} \\
\vs{-2} \\
&&\dsp{\hs{4.3} + M (R-2M)^2 u^t \big[ - M\eta \lh R(3R - 4M) + a^2  \rh } \\
&&\dsp{\hs{12}  + a\ve \lh R(2R^2 + MR - 4M^2) + a^2 M  \rh \big]} \\
\vs{-2} \\
&&\dsp{\hs{4.3} +(R-2M)^2  u^{\vf} \big[ aM \eta \lh R (2R^2 + MR - 4M^2) + a^2 M \rh} \\
&&\dsp{\hs{6.5} + \ve \lh -R^4 (R-2M)^2 + a^2 R (-3R^3 + M^2 R + 4M^3) - a^4 M^2\rh \big] \big\}}.
 \ea
\ee


\np

\bibliographystyle{newutphys}
\bibliography{bibliography_spinletter}

\end{document}